\documentclass[12pt,onecolumn]{IEEEtran}
\usepackage{bbm}
\usepackage{tipa}
\usepackage{mathrsfs}
\usepackage[noend]{algpseudocode}
\usepackage{algorithmicx,algorithm}
\usepackage[dvips]{graphicx}
\usepackage{amsfonts}
\usepackage{amssymb}
\usepackage{amsmath}
\usepackage{mathrsfs}
\usepackage{verbatim}
\usepackage{subfigure}
\usepackage{balance}
\usepackage{booktabs}
\usepackage{fancybox}
\usepackage{bm}
\usepackage{pifont}
\usepackage{extarrows}
\usepackage{cite}
\usepackage{cases}
\usepackage{rotating}
\usepackage{xcolor}

\newtheorem{proof}{Proof}

\begin{document}
\bibliographystyle{IEEEtran}
\baselineskip 4.2ex

\title{Massive Random Access with Sporadic Short Packets: Joint Active User Detection and Channel Estimation via Sequential Message Passing
\thanks{}
\author{Jia-Cheng Jiang and Hui-Ming Wang, \emph{Senior Member}, \emph{IEEE}\hspace{0.02in}}
\thanks{Jia-Cheng Jiang and Hui-Ming Wang are with the
School of Electronics and Information Engineering, and also with the Ministry
of Education Key Lab for Intelligent Networks and Network Security, Xi'an Jiaotong
University, Xi'an, 710049, Shaanxi, P. R. China. Email: {\tt j1143484496b@stu.xjtu.edu.cn; xjbswhm@gmail.com.}}
\thanks{}
}
\maketitle

\begin{abstract}
This paper considers an uplink massive machine-type communication (mMTC) scenario, where a large number of user devices are connected to a base station (BS). A novel grant-free massive random access (MRA) strategy is proposed, considering both the sporadic user traffic and short packet features. Specifically, the notions of active detection time (ADT) and active detection period (ADP) are introduced so that active user detection can be performed multiple times within one coherence time. By taking sporadic user traffic and short packet features into consideration, we model the joint active user detection and channel estimation issue into a dynamic compressive sensing (CS) problem with the underlying sparse signals exhibiting substantial temporal correlation. This paper builds a probabilistic model to capture the temporal structure and establishes a corresponding factor graph. A novel sequential approximate message passing (S-AMP) algorithm is designed to sequentially perform inference and recover sparse signal from one ADT to the next. The Bayes active user detector and the corresponding channel estimator are then derived. Numerical results show that the proposed S-AMP algorithm enhances active user detection and channel estimation performances over competing algorithms under our scenario.
\end{abstract}

\section{Introduction}
Recently, the development of the 5G cellular communication systems drives a number of newly emerging use cases. This leads to the key requirements
to support massive machine-type communications (mMTC), providing connectivity for millions of devices that perform machine-centric tasks
such as environment sensing, surveillance, control and event detection \cite{YuOn2017}. Different from the conventional human-centric
communications network, such as 4G LTE, aiming for a high data rate using large packet
sizes, the core mission in the mMTC scenario changes into the uplink access with a low transmission rate. The common features of the typical mMTC application scenarios are: massive number of user devices,
sporadic user activity and small data packets \cite{BockelmannMassive2016}. Such completely different assumptions compared with those in human-centric communication systems trigger a completely different set of technologies.

To be more specific, the non-orthogonal medium access and grant-free access control have been considered to provide access for a massive
number of devices in the uplink \cite{LiuMassive12018, LiuSparse2018}. Typically, the massive number of user devices makes it impossible to
assign orthogonal pilot sequences to all potential user devices. Hence, the non-orthogonal sequences are considered in the preamble design to
enable a certain degree of temporal resource overloading. For the access control, the traditional strategies are
grant-based, where devices access the network with a prior scheduling assignment, which requires good predictions of the
uplink requests, as well as additional control signaling or message exchanges to facilitate the granting of resources
\cite{BockelmannMassive2016}. When the base station (BS) detects multiple user access requests simultaneously, the affected user devices will be arranged to
restart the access procedure after a time expire \cite{SenelGrant-Free2018}. However, for the case of mMTC, grant-based access design
is difficult and potentially inefficient to support massive connectivity to access the network. Therefore, the promising access control pattern
in mMTC applications is the grant-free random access scheme, where each active device directly transmits its unique preamble sequence to the
BS without waiting for any permission, which results in a low control overhead.

However, non-orthogonal sequence based
grant-free access often suffers from collisions,
namely, multiple users access concurrently, and thus cannot be successfully detected and decoded \cite{BockelmannMassive2016}. Therefore, joint active user detection and
channel estimation becomes a critical issue for mMTC applications.
To address such a problem, the sparse feature of mMTC can be taken into consideration. Due to the low-rate feature of the devices in mMTC scenario, it is always the case that most devices sleep most of the time for energy efficiency and are only activated when
triggered by external events. This causes that the traffic pattern for each user device is sporadic and partly unpredictable with only a small subset of users being active concurrently. From a physical layer perspective, this situation leads to a sparse recovery problem in mMTC.
Compressive sensing (CS) technology is therefore promising to provide an advanced collision resolution to serve massive user devices, where randomly generated non-orthogonal pilot sequences are assigned to user devices
\cite{Schepker2013Exploiting, Wunder2015Compressive, Xu2015Active, LiuMassive12018}. Especially, the approximate message passing (AMP)
algorithm \cite{DonohoMessage} has attracted high attentions to efficiently cope with the challenge of joint active user detection and channel
estimation under the massive random access (MRA) scenario in \cite{LiuMassive12018, ChenSparse2018, SenelGrant-Free2018, HannakJoint2015}. The AMP
framework can achieve a high efficiency, and the performance of AMP can be evaluated via the so-called state evolution equation. The authors
in \cite{LiuMassive12018, ChenSparse2018} have shown that the prior statistic knowledge of the wireless channel can be exploited by modifying the denoiser function in the AMP framework to enhance the detection performance.

These works exhibit efficiencies of the CS technology to provide joint active user detection and channel estimation strategies with large number of devices and sporadic user traffic, which are the two primary standard literature assumptions of the mMTC scenario. However, the designs of the existing algorithms did not further take the short packet feature into account.
Typically, the average length of packets potentially
goes down to a few bytes in mMTC scenario \cite{BockelmannMassive2016}, leading to that the transmission duration of one user device is far shorter
than that of the traditional human-centric communications \cite{BockelmannMassive2016, YangWireless2018}. And, it is often the case that the
geographical locations of devices change negligibly, which results in the insignificant fluctuation of user channel coefficients and therefore
a long coherence time. Further, as discussed, the event-driven traffic makes the access patterns unpredictable and sporadic.
These scenarios and features will give rise to the facts that the user devices can potentially exist multiple state switches, i.e., on-off, within one coherence time duration, and the times of access are totally random and unpredictable.

However, the existing algorithms, such as \cite{LiuMassive12018, LiuSparse2018}, assume that all the user devices maintain active or inactive throughout the whole coherent time duration, which does not fully conform the above features of mMTC. Specifically, user devices are considered synchronized, and the joint activity detection and channel estimation operations are implemented once within a coherence block, and data transmissions of the active user devices continue until the start of the next coherent time block. Further, the \emph{independent} block-fading channel model is considered, which assumes that all the channels follow independent quasi-static flat fading within a block of coherence time. Such conditions are too restrictive for mMTC with short packets and unpredictable traffics.

In this work, we design a new scheme to achieve grant-free MRA with considering the features of both the sporadic traffic and the short packet. To be more specific, we define active detection time (ADT) as the time point for user activity detection and channel estimation. The interval between the two adjacent ADTs is denoted as the active detection period (ADP), which can be divided into two phases. In the first phase, all active user devices transmit unique non-orthogonal pilot sequences to enable the active
user detection and channel estimation. And in the second phase, data sequences with low rates will be transmitted. Note that although we still
assume the synchronous user access, the duration of an ADP is much shorter than the coherence time, so that the user requests could be responded in a timely manner. In addition, the user devices are permitted to access multiple times within one coherence time. These features fit the sporadic traffic and the short packet features well compared with the traditional synchronous access schemes mentioned above, such as \cite{LiuMassive12018, LiuSparse2018}.


Taking all the above features into consideration, in this paper, we model the joint user detection and channel estimation issue into a dynamic CS problem. Specifically, we denote the underlying time-varying sparse signal using \emph{access state sparse vector}, which exhibits substantial temporal correlation in two aspects. First, the active user indicator, which is the support vector of sparse signal, varies correlatively with ADTs. Second, the channel coefficients of user devices, which is the amplitude of the sparse signal, changes smoothly with ADTs.

In mMTC applications, the dynamic CS algorithms have been considered for multi-user detection \cite{Dynamic2016, Efficient2017}, where the temporal correlation of the user support between adjacent time step is exploited. Particularly, the authors in \cite{Efficient2017} further utilized the quality of the prior-information support set. Some related works in the signal processing literature have considered to solve the dynamic CS problem. Algorithms in \cite{Angelosante20091, Vaswani2010} are inspired by convex relaxation. On the other hand, the authors in  \cite{Angelosante20092, ZinielDynamic2013} consider a Bayesian framework. Specifically, \cite{Angelosante20092} blends elements of Bayesian models with more traditional CS through convex relaxation and greedy methods, while in \cite{ZinielDynamic2013}, the authors consider a full Bayesian framework but the algorithm is heuristic and lack of theoretical measure rule.

To the best of our knowledge, our paper is the first to consider the temporal correlations of both the channel coefficient and the active user indicator in the massive connectivity literature. In this paper, we utilize the temporal correlations for both user support and user channel, and design a novel AMP-based method for solving our dynamic CS problem under the Bayesian framework. The performance of our proposed algorithm can be predicted by state evolution equation. To passing the message from one ADT to the next, we design a distribution approximation strategy based on moment-matching, which is optimal under a typical measure rule. Our contributions can be summarized as follow.

\begin{itemize}
\item
We build a probabilistic model to reflect the temporal structure under the proposed MRA strategy. We establish a specific factor graph model in our scenario and provide the corresponding message passing schedule to implement the message passing algorithm under our graph model.
\item
We propose a novel sequential message passing algorithm to recursively recover the access state sparse vector. Specifically, we utilize the AMP framework based on the historical knowledge-aided prior. We derive the historical knowledge-aided prior based on the moment matching equations, which is optimal in the perspective of  Kullback-Leibler (KL)-divergence. The state evolution analysis is provided, indicating that the historical knowledge benefits the AMP framework in our scenario.

\item
We derive the active user detection and channel estimation strategy, which can be performed in each ADT after executing the proposed message passing algorithm. Specifically, the LLR test with Bayes criterion is considered and the channel estimation can be derived directly based on the recovered sparse vector.

 \end{itemize}

\emph{Notation}:
Throughout this paper, scalars are denoted by lower-case letters, vectors by bold-face lower-case letters, and matrices by bold-face upper-case
letters. For a matrix $\boldsymbol A$, $\boldsymbol A^{T}$ and $\boldsymbol A^{H}$ denote its transpose and conjugate transpose, respectively. The ${\rm
Pr}\{\cdot\}$ returns the probability mass and $p(\cdot)$ returns the probability density. $\{\cdot\}_{a}^{b}$ returns the collection of variables from index $a$ to $b$. The distribution of a circularly symmetric
complex Gaussian random vector $\boldsymbol x$ with mean $\boldsymbol \mu$ and covariance matrix $\boldsymbol\Sigma$ is denoted by
$\mathcal{CN}(\boldsymbol x;\boldsymbol \mu, \boldsymbol\Sigma)$. Finally, $\mathfrak R(\cdot)$ returns the real part of the variable.

\section{System Model}
In this paper, we consider the uplink of a mMTC scenario with one BS located at the center of the cellular and $N$ devices located
randomly in a coverage area. For simplicity, we assume that the BS as well as each device is equipped with a single antenna. User device $n$ is assigned a unique pilot sequence of length $L$, denoted as $\boldsymbol s_n \in
\mathbb C^{L}\triangleq [s_{1n}, s_{2n},\dots, s_{Ln}]^{T}$. Since we
are interested in the scenario that the number of potential user devices is much larger than the length of pilot sequence, i.e., $N\gg L$, the
non-orthogonal pilot sequences are assigned to user devices. We further assume that the pilot sequence is generated according to an i.i.d. complex
Gaussian distribution with zero mean and variance $1/L$ such that each sequence has a unit power \cite{LiuMassive12018, ChenSparse2018}.

\subsection{Access Strategy}
\begin{figure}[!t]
\centering
\includegraphics[width=3.5in]{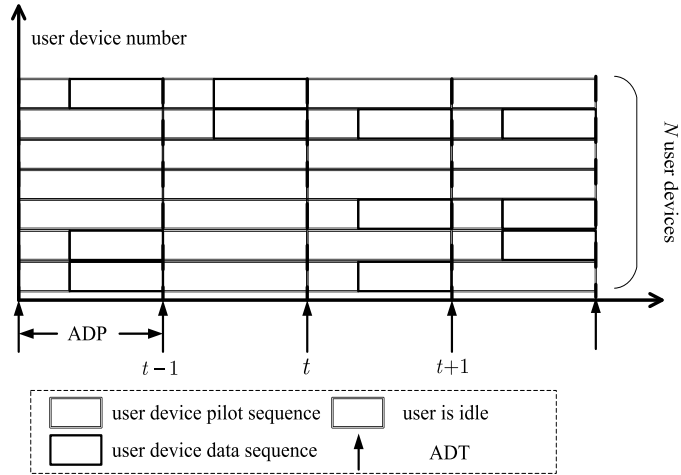}
\caption{Sporadic and short packet transmission pattern for mMTC.}
\label{system}
\end{figure}

In this work, we define the concepts of ADT and ADP, which can be seen in Fig. \ref{system}. We consider a synchronous access strategy, and each user device is permitted to choose whether or not to access the network in each ADT. For the sporadic nature of user traffic, in the $t$th ADT, there is only a subset of the users that are active, and the other users are idle. We denote $a_n^{(t)}$ as the active user indicator in $t$th ADT, with $a_n^{(t)} = 0$ or $1$
indicating the user is idle or active, respectively. The received signal in the $t$th ADT at the BS can be modeled as
\begin{equation}
\boldsymbol y^{(t)} = \sum\limits_{n = 1}^{N}a_n^{(t)}\boldsymbol s_nh_n^{(t)}+\boldsymbol w^{(t)},\label{sy1}
\end{equation}
where $h_n^{(t)}\in\mathbb C$ is the channel coefficient between user $n$ and BS, and $\boldsymbol w^{(t)}\in\mathbb C^{L}$ is the corresponding complex Gaussian noise vector with each element $w_n^{(t)}\sim
\mathcal{CN}(w_n^{(t)};0,\sigma_w^2)$. Note that the user transmit power $p_n$ is absorbed into the channel coefficient for concise. We define $x_n^{(t)} \triangleq a_n^{(t)}h_n^{(t)}$, and the vector $\boldsymbol x^{(t)}
\triangleq [x_1^{(t)}, x_2^{(t)},\dots, x_N^{(t)}]^{T}\in
\mathbb C^{N}$ forms a sparse time vector in each ADT, which is denoted as access state sparse vector.
As a consequence, the system model in (\ref{sy1}) can be restated as
\begin{equation}
\boldsymbol y^{(t)} = \boldsymbol S\boldsymbol x^{(t)}+\boldsymbol w^{(t)}.\label{sy2}
\end{equation}
where $\boldsymbol S\triangleq [\boldsymbol s_1,\dots, \boldsymbol s_N]\in\mathbb C^{L\times N}$. For the sake of presentation, according to (\ref{sy2}), we define $\boldsymbol a^{(t)} \triangleq [a_1^{(t)}, a_2^{(t)},\dots, a_N^{(t)}]^{T}$, $\boldsymbol h^{(t)}
\triangleq [h_1^{(t)}, h_2^{(t)},\dots, h_N^{(t)}]^{T}$. And $\boldsymbol Y \triangleq \{\boldsymbol y^{(t)}\}_{t =
1}^{T}\in\mathbb C^{L\times T}$, $\boldsymbol X \triangleq \{\boldsymbol x^{(t)}\}_{t = 1}^{T}\in\mathbb C^{N\times T}$, $\boldsymbol A\triangleq\{\boldsymbol a^{(t)}\}_{t = 1}^{t}\in\mathbb C^{N\times
T}$, $\boldsymbol H\triangleq\{\boldsymbol y^{(t)}\}_{t = 1}^{T}\in\mathbb C^{N\times T}$ are
denoted as the collections of $\boldsymbol y^{(t)}$, $\boldsymbol x^{(t)}$, $\boldsymbol a^{(t)}$, $\boldsymbol h^{(t)}$ in $T$ consecutive ADTs, respectively.

Note that for the typical scenario in mMTC, where the mobility of the user devices is negligible, the user channel coherence time could be extremely long, so that the user devices may potentially exist
multiple state switches, i.e., on-off, with unpredictable times of accesses within one coherence time duration. Although we still assume the synchronous user access strategy, the duration of an ADP is designed much shorter than the length of a coherence block, so that the user requests could be responded in a timely manner. Moveover, the user devices are permitted to access multiple times within one coherence block. Such a design fits the sporadic traffic and short packet features well compared with the access schemes in \cite{LiuMassive12018, LiuSparse2018}, where they consider that there is only one ADT in each period of coherence time. Such a design is also compatible with the case that different user devices belong to different terminal patterns. For example, when the packet length and the transmission duration for a user device are relatively short, the user device may access at one ADT and immediately disconnect from the network at the next. On the contrary, the user with a long transmission duration typically covers several consecutive ADPs. As shown in Fig. \ref{system}, there could exist \romannumeral1) multiple consecutive ADTs within one user device transmission, and \romannumeral2) multiple consecutive ADTs within
one coherence time. Obviously, the access state sparse vector $\boldsymbol x^{(t)}$ often exhibits a high degree of correlation from one ADT to the next, which reflects in two aspects. First, the active user indicator $\boldsymbol a^{(t)}$, which can be regarded as the support vector of the corresponding sparse vector $\boldsymbol x^{(t)}$, is highly correlated in the adjacent ADTs. Second, the channel coefficient $\boldsymbol h^{(t)}$, which is the amplitude of $\boldsymbol x^{(t)}$, changes smoothly with ADTs.

\subsection{Probabilistic Model}
To characterize the time-variation of the active user indicator $\boldsymbol a^{(t)}$, and the smooth evolution of the channel vector
$\boldsymbol h^{(t)}$, we consider a probabilistic model as follow. We model the change of the $n$th element of the support vector $a_n^{(t)}$
across time as a Markov chain characterized by a couple of transition probabilities, i.e., $p_n^{(10)}\triangleq{\rm Pr}\{a_n^{(t)} = 1|a_n^{(t-1)} =
0\}$ and  $p_n^{(01)}\triangleq{\rm Pr}\{a_n^{(t)} = 0|a_n^{(t-1)} = 1\}$, and $N$ users are supposed to form independent Markov chains. We
further assume that each chain is under a steady state with ${\rm Pr}\{a_n^{(t)} = 1\} = \lambda_n$ that indicates the activation probability of
each user $n$. Under this condition, the Markov chain of each user $n$ can be specified by parameters $p_n^{(01)}$ and $\lambda_n$, with the transition probability $p_n^{(10)}$ formulated as $p_n^{(10)} = \lambda_np_n^{(01)}/(1-\lambda_n)$. Especially, we assume that in each ADT the active user ratio is $\lambda$ and the ratio of users from active to idle is $p_{01}$, and the transition and activation probabilities are independent with $n$, i.e., $p_n^{(10)} = p_{10},p_n^{(01)} = p_{01}, \lambda_n = \lambda, \forall n$. Note that such an assumption captures the transition ratio and active user ratio in each ADT, and has shown its efficiency in the similar application \cite{ZinielDynamic2013}\footnote{Although it ignores some specific information for each user, e.g., the average transmission duration of each user device may be different, it has shown its efficiency in \cite{ZinielDynamic2013}}. The probabilistic
distribution that specifies the Markov chains can be given as
\begin{equation}
p(a_n^{(t)}|a_n^{(t-1)}) =
(1-p_{10})^{(1-a_n^{(t)})(1-a_n^{(t-1)})}p_{10}^{a_n^{(t)}(1-a_n^{(t-1)})}(1-p_{01})^{a_n^{(t)}a_n^{(t-1)}}p_{01}^{(1-a_n^{(t)})a_n^{(t-1)}},\label{pat}
\end{equation}
where we define $p(a_n^{(1)}|a_n^{(0)})\triangleq p(a_n^{(1)}) = (1-\lambda)^{1-a_n^{(1)}}\lambda^{a_n^{(1)}}$.

On the other hand, the smooth
evolutions of the channel coefficients for all the users can be characterized by a set of independent Gaussian Markov state-space models. Since for each user, the probabilistic distribution of the channel coefficient depends highly on the propagation environment and the geographical location changes negligibly in several channel coherence blocks, the statistical characteristics of channel
coefficient stay unchanged. Hence, we assume a steady-state Gaussian Markov processes for each user, which can be characterized by a
first order autoregressive (AR-1) model \cite{MaSparse2019, PrasadJoint2015} as
\begin{equation}
h_n^{(t)} = \eta_nh_n^{(t-1)}+u_n^{(t)},\label{h}
\end{equation}
where $\eta_n = J_0(2\pi D_nT_b)$ is the AR coefficient that controls the temporal correlation, $J_0(\cdot)$ is the zeroth order Bessel
function of the first kind, $D_n$ is the Doppler frequency of user $n$, and $T_b$ is the time duration of an ADP. We further suppose that the
Gaussian-Markov process for user $n$ is under steady state with zero mean and variance $\rho_n$, and the evolution noise $u_n^{(t)}$ is
therefore distributed as $u_n^{(t)}\sim\mathcal{CN}(u_n^{(t)}; 0,(1-\eta_n^2)\rho_n)$. Thus, the Gaussian-Markov can be specified by the following
probabilistic distribution
\begin{equation}
p(h_n^{(t)}|h_n^{(t-1)}) = \mathcal{CN}(h_n^{(t)};\eta_nh_n^{(t-1)},(1-\eta_n^2)\rho_n),\label{pht}
\end{equation}
where we define $p(h_n^{(1)}|h_n^{(0)})\triangleq p(h_n^{(1)}) = \mathcal{CN}(h_n^{(1)}; 0,\rho_n)$. Further, according to the definition of
sparse vector $\boldsymbol x^{(t)}$, we can infer that the probabilistic distribution of $x_n^{(t)}$ conditional on $h_n^{(t)}$ and $a_n^{(t)}$
is
\begin{equation}
p(x_n^{(t)}|h_n^{(t)},a_n^{(t)}) = \delta(x_n^{(t)}-h_n^{(t)}a_n^{(t)}),\label{xcha}
\end{equation}
where $\delta(\cdot)$ is the Dirac delta function. We assume that channel coefficient $h_n^{(t)}$ evolves independently from the support vector
$a_n^{(t)}$. Marginalizing out $h_n^{(t)}$ and $a_n^{(t)}$ via (\ref{pat}) and (\ref{pht}), we can obtain the marginal distribution over
$x_n^{(t)}$ as \cite{ZinielDynamic2013}
\begin{equation}
p(x_n^{(t)}) = (1-\lambda)\delta(x_n^{(t)})+\lambda\mathcal{CN}(x_n^{(t)}; 0,\rho_n).\label{px}
\end{equation}
The form of distribution (\ref{px}) is the Gaussian-Bernoulli distribution and also known as the ``spike-and-slab'' prior distribution, which
is an efficient sparsity-promoting prior with point-mass on $x_n^{(t)} = 0$ \cite{ZinielEfficient2013, ZinielDynamic2013}. The
parameter $\lambda$ controls the fraction of $x_n^{(t)}$ that is expected to be zero.

Note that the parameters for the probabilistic model can be specified by some specific information of user devices. The active ratio $\lambda$, the transition probability $p_{10}$, the channel correlation coefficient $\eta_n$, the channel variance $\rho_n$ can be specified by the user access frequency \cite{LiuMassive12018}, user transmission duration \cite{ZinielDynamic2013}, user speed \cite{PrasadJoint2015}, and the distance between user device and the BS, respectively, which are considered known for BS in this paper.

The goal for the BS is to first detect the user activities and to further estimate the corresponding channel coefficient for each active user.
This can be done by recovering the access state sparse vector $\boldsymbol x^{(t)}$ in each ADT. Since we have introduced the temporal
correlation between the user indicator (\ref{pat}) and the temporal correlation between channel coefficient (\ref{pht}), we consider
recursively recover the sparse vector $\boldsymbol x^{(t)}$. This forms a dynamic CS problem, which is different from the sparse recovery algorithms in
traditional MRA models \cite{LiuMassive12018, ChenSparse2018}, where the recovery is performed \emph{independently} with the invariant prior
distribution (\ref{px}). The proposed algorithm to recursively recover the sparse signal is called sequential approximate message passing (S-AMP). The algorithm mainly focuses on the following issues: how to recover the sparse signal in current ADT with historical knowledge and how to deliver the knowledge from the current ADT to the next.

 \section{S-AMP: Graph Representation and Schedule}
Our inference is based on the message passing framework under a specific factor graph of our system model. In this section, we specify the factor graph and design the message passing schedule in our model.

\subsection{Graph Representation}
\begin{figure}[!t]
\centering
\includegraphics[width=3.5in]{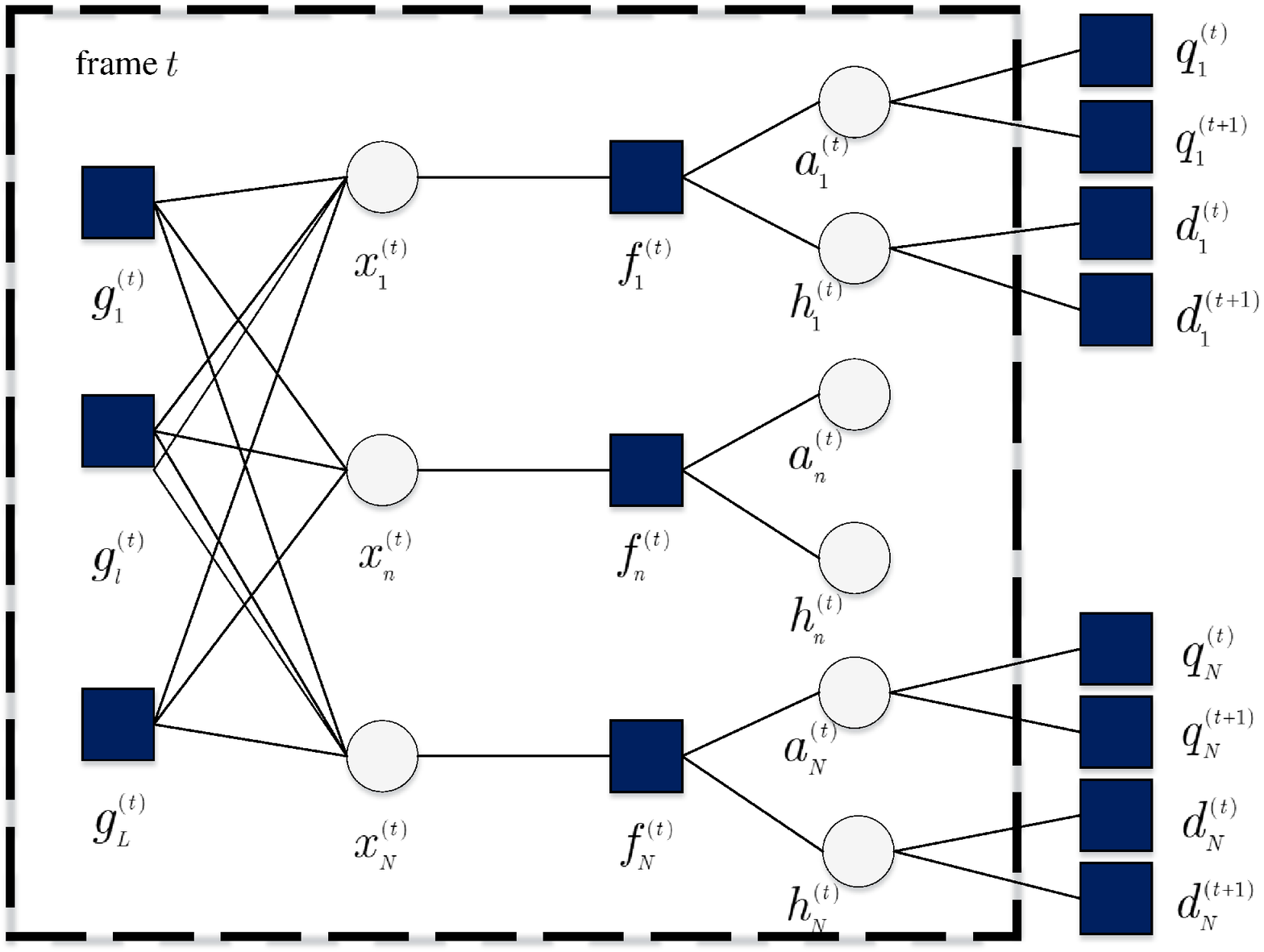}
\caption{Factor graph representation of the proposed model.}
\label{graph}
\end{figure}

The factor graph is derived based on the decompositions of a joint distribution.
Exploiting the inherent statistical structure of our model (\ref{system}), the corresponding joint distribution of sparse signals,
active indicators, channel coefficients can be decomposed as \begin{equation}
p(\boldsymbol X, \boldsymbol A, \boldsymbol H|\boldsymbol Y)  = Z^{-1}\prod_{t=1}^{T}\prod_{l=1}^{L}p(y_l^{(t)}|\boldsymbol
x^{(t)})\prod_{n=1}^{N}p(x_n^{(t)}|a_n^{(t)},h_n^{(t)})p(a_n^{(t)}|a_n^{(t-1)})p(h_n^{(t)}|h_n^{(t-1)}),\label{promodel}
\end{equation}
where $Z$ is a normalized constant, $p(y_l^{(t)}|\boldsymbol x^{(t)}) = \mathcal{CN}(y_l^{(t)};\boldsymbol s_l^{T}\boldsymbol x^{(t)},
\sigma_w^2)$ with $y_l^{(t)}$ denoted the $l$th element of $\boldsymbol y^{(t)}$ and $\boldsymbol s_l^{T}$ denoted the $l$th row of the pilot
sequence matrix $\boldsymbol S$. We then give notations of the factor nodes within (\ref{promodel}) as $g_l^{(t)}(\boldsymbol x^{(t)})
\triangleq  p(y_l^{(t)}|\boldsymbol x^{(t)})$, $f_n^{(t)}(x_n^{(t)},a_n^{(t)},h_n^{(t)}) \triangleq  p(x_n^{(t)}|a_n^{(t)},h_n^{(t)})$,
$q_n^{(t)}(a_n^{(t)},a_n^{(t-1)}) \triangleq  p(a_n^{(t)}|a_n^{(t-1)})$ and $d_n^{(t)}(h_n^{(t)},h_n^{(t-1)}) \triangleq
p(h_n^{(t)}|h_n^{(t-1)})$. Note that the variable node for each observed received signal $y_l^{(t)}$ is absorbed into the factor node
$g_l^{(t)}(\boldsymbol x^{(t)})$. Then, the associated factor graph is shown in Fig. \ref{graph}

\subsection{Scheduling the Message Passing}
From Fig. \ref{graph}, we observe that all of the variables related in the $t$th ADT can be arranged on a plane, which is referred as a
``frame''. The connections between the neighboring frames are established by the temporal correlated variables $a_n^{(t)}$, $h_n^{(t)}$ and
their corresponding factor nodes $q_n^{(t)}$, $d_n^{(t)}$ for all $n$. We can observe in Fig. \ref{graph} that the temporal correlation between
variable nodes $a_n^{(t)}$ and $h_n^{(t)}$ in each $t$th ADT brings additional loops compared with the generic AMP algprithms in the traditional MRA models \cite{LiuMassive12018, ChenSparse2018}. Therefore, the specific implementing schedule for the message passing within our proposed graph model has to be designed.

The designed message passing schedule from the $t$th frame to the $(t+1)$th frame can be divided into two distinct parts. In the first part, the algorithm focuses mainly on passing the messages within the $t$th frame with given input messages from the $(t-1)$th frame. In the second part, the algorithm focuses mainly on passing the messages from the $t$th frame into the next. Specifically, for the first part of schedule, the input messages that provide current beliefs of temporal correlated variables $a_n^{(t)}$ and $h_n^{(t)}$ are delivered to $x_n^{(t)}$. Then, the node $x_n^{(t)}$ updates the message with given received signals available in the $t$th frame, i.e., $\{\boldsymbol y^{(t)}\}_{t = 1}^{t}$. Finally, the updated messages are output from the node $x_n^{(t)}$. For the second part, the node $x_n^{(t)}$ propagates the messages providing the updated beliefs to $a_n^{(t)}$ and
$h_n^{(t)}$. Then, such messages are further propagated to the next $(t+1)$th frame as the input messages that carry the beliefs of $a_n^{(t+1)}$ and $h_n^{(t+1)}$.

We note that in the first part of the schedule, the algorithm includes messages exchange between nodes $g_l^{(t)}$ and $x_n^{(t)}$. The matrix $\boldsymbol S$ couples a sequence of nodes $\{x_n^{(t)}\}_{n = 1}^N$ into each $g_l^{(t)}$, leading to several loops between nodes $g_l^{(t)}$ and $x_n^{(t)}$. Hence, the corresponding messages passing algorithm in the first part of the schedule is required to be executed iteratively. On the contrary, we observe that in the second part of the schedule, the messages are propagated only in one direction. Namely, once the messages are passed from the current frame to the next, they will not feed back to re-update the belief in the current frame. As a consequence, we note that the convergence condition of the proposed algorithm is same as the AMP algorithms \cite{DonohoMessage}. For large but finite-sized i.i.d. Gaussian matrix $\boldsymbol S$, the AMP performance is shown to be close to Bayes-optimal \cite{Rush2016}. Moreover, the AMP framework has been shown to perform extremely well in a number of applications in the mMTC literature, such as \cite{LiuMassive12018, ChenSparse2018}.
In the following sections, we consider the design of the concrete S-AMP algorithm based on such a schedule.

\section{S-AMP: AMP Based on Historical Knowledge-Aided Prior}
In this section, we focus on the message passing algorithm under the first part of the schedule, where it absorbs the messages from the $(t-1)$th frame and updates the messages with the available received signal in the $t$th frame. We assume that the messages from $(t-1)$th ADT to the $t$th are known, and we will show that the proposed algorithm in this part based on such messages is equivalent to the generic AMP algorithm based on the historical knowledge-aided prior.
\subsection{AMP Based on Historical Knowledge-Aided Prior}
The first part of the schedule can be further partitioned into three distinct steps, which are denoted as ``into'', ``within'', and ``out'' step, respectively. Specifically, in each ADT, the ``into'' step involves the passing messages that provide current beliefs of temporal correlated variables $a_n^{(t)}$ and $h_n^{(t)}$ and forms the historical knowledge-aided prior of $x_n^{(t)}$ for each user device $n$. The ``within'' step utilizes the prior, together with the observations in the $t$th ADT to generate the posterior estimation of sparse signal $\boldsymbol x^{(t)}$, aiming to achieve minimum mean square error (MMSE). The ``out'' step feeds back the updated messages to every node $x_n^{(t)}$, which will be utilized in the second part of the schedule.
\subsubsection{``into'' step}\label{into} Since we concern a filtering-like framework, no information will be conveyed
from the unexperienced frame $(t+1)$ in current $t$th frame. Hence, it is equivalent to disconnecting the links between factor nodes $q_n^{(t+1)}$,
$d_n^{(t+1)}$ and variable nodes $a_n^{(t)}$, $h_n^{(t)}$ in this step. Therefore, the message $\nu_{f_n^{(t)}\to x_n^{(t)}}$ can be derived as
\begin{equation}
\nu_{f_n^{(t)}\to x_n^{(t)}}(x_n^{(t)})\propto\sum\limits_{a_n^{(t)} = \{0,1\}}\int_{h_n^{(t)}}f_n^{(t)}(x_n^{(t)}, a_n^{(t)},
h_n^{(t)})\cdot\nu_{a_n^{(t)}\to f_n^{(t)}}(a_n^{(t)})\cdot\nu_{h_n^{(t)}\to f_n^{(t)}}(h_n^{(t)}),\label{lo_prx}
\end{equation}
where $\nu_{a_n^{(t)}\to f_n^{(t)}}(a_n^{(t)}) = \nu_{q_n^{(t)}\to a_n^{(t)}}(a_n^{(t)})$ and $\nu_{h_n^{(t)}\to f_n^{(t)}}(h_n^{(t)}) =
\nu_{d_n^{(t)}\to h_n^{(t)}}(h_n^{(t)})$ because there is only one edge between both $a_n^{(t)}$ and $q_n^{(t)}$ as well as $h_n^{(t)}$ and
$d_n^{(t)}$. The messages $\nu_{q_n^{(t)}\to a_n^{(t)}}(a_n^{(t)})$ and $\nu_{d_n^{(t)}\to h_n^{(t)}}(h_n^{(t)})$ provide current beliefs of variables $a_n^{(t)}$ and $h_n^{(t)}$, which absorb the messages from $(t-1)$th frame, carrying historical knowledge. Since the initial forms
of messages $\nu_{q_n^{(1)}\to a_n^{(1)}}(a_n^{(1)})$ and $\nu_{d_n^{(1)}\to h_n^{(1)}}(h_n^{(1)})$ are Bernoulli and Gaussian distribution, respectively, we assume that
\begin{equation}
\nu_{q_n^{(t)}\to a_n^{(t)}}(a_n^{(t)} = 1) \triangleq \hat{\pi}_{n,t},\quad\nu_{d_n^{(t)}\to h_n^{(t)}}(h_n^{(t)}) \triangleq
\mathcal{CN}(h_n^{(t)};\hat{\xi}_{n,t},\hat{\psi}_{n,t}),\label{dth}
\end{equation}
where $\hat{\pi}_{n,t}$, $\hat{\xi}_{n,t}$ and
$\hat{\psi}_{n,t}$ are the parameters of the prior and assumed known in this part. Particularly, in the first frame, we have $\hat{\pi}_{n,1} = \lambda$ and $\hat{\xi}_{n,1} = 0$ and $\hat{\psi}_{n,1} = \rho_n$ for each $n$. As a
consequence, the prior of the node $x_n^{(t)}$ in (\ref{lo_prx}) can be formulated as
\begin{equation}
\nu_{f_n^{(t)}\to x_n^{(t)}}(x_n^{(t)}) =
(1-\hat{\pi}_{n,t})\delta(x_n^{(t)})+\hat{\pi}_{n,t}\mathcal{CN}(x_n^{(t)};\hat{\xi}_{n,t},\hat{\psi}_{n,t}).\label{lopx}
\end{equation}
Comparing this historical knowledge-aided prior (\ref{lopx}) of $x_n^{(t)}$ with (\ref{px}), we find that both of them are
`spike-and-slab'' prior distributions, which are sparsity-promoting. The difference is that the updated parameters $\hat{\pi}_{n,t}$, $\hat{\xi}_{n,t}$ and
$\hat{\psi}_{n,t}$ in (\ref{lopx}) contain historical knowledge
about the received signals $\{\boldsymbol y^{(t)}\}_{t = 1}^{t-1}$.
\subsubsection{``within'' step}:\label{within} In this step, our task is to utilize the historical knowledge-aided prior (\ref{lopx}), together with the observations $\boldsymbol y^{(t)}$ in current $t$th ADT, to generate the posterior estimation of sparse signal $\boldsymbol x^{(t)}$, aiming to achieve MMSE. The main difficulty for this problem is that the matrix $\boldsymbol S$ mixes the coefficients of $\boldsymbol x^{(t)}$ into $\boldsymbol y^{(t)}$. Fortunately, in large system limit, i.e. $L,N\to\infty$ with $L/N$ fixed, such a vector-valued estimation problem can be efficiently solved via the generic AMP framework \cite{SchniterTurbo2010}, and an estimate of $\boldsymbol x^{(t)}$ based on $\boldsymbol y^{(t)}$ that minimizes the mean-squared error (MSE) could be obtained. For concise, we drop the superscript of index $t$. The generic AMP initializes $\mu_{n}^{0} = 0$, $z_{l}^{0} = y_l$, and $c^{0}\gg\sigma_w^2$ for all $n$ and $l$, and then iterates the following equations for $i$th iteration,
\begin{align}
&\phi_n^{i} = \sum\nolimits_{l = 1}^{L}S_{ln}^*z_{l}^i+\mu_{n}^{i},\label{amps}\\
&\mu_{n}^{i+1} = F_n(\phi_n^{i},c^{i}),\quad v_{n}^{i+1} = G_n(\phi_n^{i},c^{i}),\label{mmseden}\\
&z_{l}^{i+1}= y_l-\sum\nolimits_{n = 1}^{N}S_{ln}\mu_{n}^{i+1}+\frac{z_{l}^{i}}{L}\sum\nolimits_{n = 1}^NF_n'(\phi_n^{i},c^{i}),\\
&c^{i+1}=\sigma_w^2+
1/L\sum\nolimits_{n = 1}^{N}v_{n}^{i+1},\label{ampe}
\end{align}
where $F_n'(\phi_n^{i},c^{i})\triangleq \frac{\partial F_n(\phi_n^{i},c^{i})}{\partial \phi_n^{i}}$ is the first derivative of function $F_n(\phi_n^{i},c^{i})$ with respect to $\phi_n^{i}$. Using
(\ref{lopx}), (\ref{proG}) and together with the definitions, the specific expressions of above functions are given as
\begin{align}
&F_n(\phi_n^{i},c^{i})=
(1+\gamma_n(\phi_n^{i},c^{i}))^{-1}\left(\frac{\hat{\psi}_{n}\phi_n^{i}+\hat{\xi}_{n}c^{i}}
{\hat{\psi}_{n}+c^{i}}\right),\quad F_n'(\phi_n^{i},c^{i})=\frac{1}{c^{i}}G_n(\phi_n^{i},c^{i}),\label{mmse}\\
&G_n(\phi_n^{i},c^{i})=(1+\gamma_n(\phi_n^{i},c^{i}))^{-1}\left(\frac{\hat{\psi}_{n}c^{i}}
{\hat{\psi}_{n}+c^{i}}\right)+\gamma_n(\phi_n^{i},c^{i})|F_n(\phi_n^{i},c^{i})|^2,
\end{align}
where
\begin{equation}
\gamma_n(\phi_n^{i},c^{i})\triangleq\left(\frac{1-\hat{\pi}_{n}}{\hat{\pi}_{n}}\right)
\left(\frac{\hat{\psi}_{n}+c^{i}}{c^{i}}\right){\rm exp}\left(-\left[\frac{\hat{\psi}_{n}|\phi_n^{i}|^2+2\mathfrak
R(\hat{\xi}_{n}^*c^{i}\phi_n^{i})-c^{i}|\hat{\xi}_{n}|^2}{c^{i}(\hat{\psi}_{n}+c^{i})}\right]\right).\label{gamma}
\end{equation}

After the convergence of the generic AMP algorithm (\ref{amps})-(\ref{ampe}), the posterior estimation of the sparse vector $\boldsymbol x$ is given by $ \boldsymbol {\hat x} = \boldsymbol\mu^{I}$, where we denote $\boldsymbol\mu^{I}\triangleq [\mu^{I}_1, \mu^{I}_2,\dots, \mu^{I}_N]^{T}\in\mathbb C^{N}$ and the index $I$ indicates the maximal number of AMP iteration times.

\subsubsection{``out'' step}So far, we have considered the messages exchange between nodes $\{x_n\}_{n = 1}^{N}$ and $\{g_l\}_{l = 1}^{L}$ with the historical knowledge-aided prior
(\ref{lopx}). Next, we derive the output message from
$x_n$, i.e., the message passing from $x_n$ to $f_n$, which will be used in the second part of our schedule. In the large system limit, it is reasonable to regard the message $\nu_{g_l\to x_n}^{i}(x_n)$ as Gaussian because of the Berry-Esseen central limit theorem \cite{SchniterTurbo2010}. This Gaussian quantity in each $i$th iteration can be
parameterized by the mean $\mu_{nl}^{i}$ and variance $v_{nl}^{i}$ of the message $\nu_{x_n\to g_l}^{i}(x_n)$. According to \cite{KschischangFactor2001}, the message $\nu_{g_l\to x_n}^{i}(x_n)$ takes the form as $\nu_{g_l\to x_n}^{i}(x_n) = \mathcal{CN}(S_{ln}x_n;z_{ln}^{i},c_{ln}^{i}),\label{gtx}$, where we define $z_{ln}^{i}\triangleq y_l-\sum\nolimits_{q\neq n}S_{lq}\mu_{ql}^{i}$ and $c_{ln}^{i}\triangleq\sigma_w^2+\sum\nolimits_{q\neq
n}|S_{lq}|^2v_{ql}^{i}$. By applying the fact
\begin{equation}
\prod_q\mathcal{CN}(x;\mu_q,v_q)\propto \mathcal{CN}(x;\frac{\sum_q\mu_q/v_q}{\sum_q v_q^{-1}},\frac{1}{\sum_q v_q^{-1}}),\label{proG}
\end{equation}
and together with the sum product algorithm, we can obtain
\begin{equation}
\nu_{x_n\to f_n}(x_n) = \prod\nolimits_{l}\nu_{g_l\to x_n}^{i}(x_n) =\mathcal{CN}(x_n;\sum\nolimits_{l}S_{ln}^*z_{ln}^{i},c_n^i).\label{xtf1}
\end{equation}
For further derivation, we utilize the general assumptions of the generic AMP framework that $z_{ln}^{i} = z_l^{i}+\delta z_{ln}^i+\mathcal O(1/N)$ and $\mu_{nl}^{i} =
\mu_n^{i}+\delta\mu_{nl}^i+\mathcal O(1/N)$, then the mean $z_{ln}^{i}$ can be rewritten as
\begin{equation}
z_{ln}^{i}= y_l-\sum\nolimits_{n}S_{ln}\mu_{nl}^{i}+S_{ln}\mu_{nl}^{i}+\mathcal O(1/N) = z_l^{i}+S_{ln}\mu_{n}^{i}+\mathcal O(1/N).\label{zn}
\end{equation}
We note that the term $\delta\mu_{nl}^i$ is absorbed into $\mathcal O(1/N)$, since $S_{ln}$ is also a $\mathcal O(1/N)$ term. Then,
substituting (\ref{zn}) into (\ref{xtf1}), after convergence, the message $\nu_{x_n\to f_n}(x_n)$  yields
\begin{equation}
\nu_{x_n\to f_n}(x_n) = \mathcal{CN}(x_n;\sum\nolimits_{l}S_{ln}^*z_{l}^{I}+\mu_{n}^{I},c^I) =\mathcal{CN}(x_n;\phi_n^I,c^I),\label{likelihood}
\end{equation}
where we have utilized the approximation $c_n^i\approx c^i$ \cite{SchniterTurbo2010}. By utilizing sum-product algorithm \cite{KschischangFactor2001}, the posterior distribution of variable $x_n$ can be obtained by multiplying the
messages from all directions to node $x_n$, i.e., $\nu_{x_n\to f_n}(x_n)$ and $\nu_{f_n\to x_n}(x_n)$. Since the message $\nu_{f_n\to
x_n}(x_n)$ is considered as the local prior for $x_n$, we imply that $\nu_{x_n\to f_n}(x_n)$ can be regarded as the approximate likelihood function of
$x_n$.

\subsection{State Evolution Analysis}
One remarkable property of the AMP framework is that the performance of sparse vector recovery can be measured by the state evolution function when the entries of the sensing matrix generated from i.i.d. Gaussian distribution \cite{BayatiDynamics2011}. In the case that the Bernoulli-Gaussian prior with the form (\ref{lopx}) is the exact prior distribution and the MMSE denoiser (\ref{mmse}) is utilized, the equation (\ref{ampe}) is exactly the state evolution function in the large system limit . For further analysis, we consider a more general form of the state evolution that applies to any arbitrary denoiser $f_{\theta_n}(\cdot, \theta_n)$ with $\theta_n$ denoted as the corresponding parameter set for each user device $n$. The general state evolution function of AMP framework is given by
\begin{equation}
c^{i+1}=\sigma_w^2+
\frac{N}{L}\mathbb E[|f_{\Theta}(X+\sqrt{c^{i}}V,\Theta)-X|^2],\label{ge_st}
\end{equation}
where the $X$, $V$, and $\Theta$ are random variables with $X$ following $p_{X|\Theta}$, $V\sim\mathcal{CN}(0,1)$, and the probability distribution of $\Theta$ denoted as $p_{\Theta}$. The expectation are taken over $X$, $V$ and $\Theta$. The state evolution equation (\ref{ge_st}) predicts the state of AMP accurately in the large system limit with the condition that the empirical distribution of $\{\theta_n\}_{n = 1}^{N}$ and $\{x_n\}_{n = 1}^{N}$ converge to the probability measure $p_{\Theta, X}$.

Further, we denote a random variable $\Phi$, and we can observe from (\ref{likelihood}) that after the convergence of AMP, the messages passing from each node $x_n$ to $f_n$ are in the form of $N$ independent Gaussian distributions. Thus, the sequence of signals $\{\phi_n\}_{n = 1}^{N}$ can be regarded as $N$ independent samples of the random variable $\Phi = X+\sqrt{c^{I}}V$, which can be interpreted as a Gaussian noise-corrupted version of $X$ with noise level $c^{I}$. Therefore, in $t$th ADT, the set of random variable $\Theta^{t}$ in (\ref{ge_st}) can be defined as $\Theta^{(t)}\triangleq\{\Gamma, \{\Phi^{(t)}\}_{t = 1}^{t-1}\}$ with a sequence of realizations $\{\theta_n^{(t)}\}_{n = 1}^{N}$, where $\theta_n^{(t)}\triangleq\{\eta_n, \rho_n, \{\phi_n^{(t)}\}_{t = 1}^{t-1}\}$, and we assume that the empirical distribution of $\{\eta_n\}_{n = 1}^{N}$ and $\{\rho_n\}_{n = 1}^{N}$ converge to the probability measure probability measure $p_{\Gamma}$.

As a consequence, in the $t$th ADT, the denoiser $f_{\theta_n}(\cdot, \theta_n)$ for each user device $n$  is designed to recover the random variable $X^{(t)}$ from its noisy version $\Phi^{(t)} = X^{(t)}+\sqrt{c_t}V$, with $X^{(t)}\sim p_{X^{(t)}|\Theta^{(t)}}(x_n^{(t)}|\theta_n^{(t)})$, and the term
$\mathbb E[|f_{\Theta^{(t)}}(X^{(t)}+\sqrt{c_t}V,\Theta)-X|^2]$ based on the state evolution equation (\ref{ge_st}) can be interpreted as the MSE of the denoiser in each iteration with given $c_t$. Note that for concise, we have omitted the superscript $i$ that will serve to keep track of the multiple iterations. Obviously, the optimal denoiser that achieves MMSE for each user is given by the expectation of the corresponding posterior distribution $p(x_n^{(t)}|\phi_n^{(t)},\theta_n^{(t)};c_t)$. In this condition, the MSE with respect to $c_t$ can be formulated as
\begin{equation}
M(c_t) = \mathbb E[{\rm Var}(X^{(t)}|\Phi^{(t)},\Theta^{(t)})],\label{ac_mse}
\end{equation}
where the ${\rm Var}(X^{(t)}|\Phi^{(t)},\Theta^{(t)})$ is the conditional variance of the posterior distribution $p_{X^{(t)}|\Phi^{(t)},\Theta^{(t)}}$ with given $\Phi^{(t)}$ and $\Theta^{(t)}$ and the expectation is taken over both $\Phi^{(t)}$ and $\Theta^{(t)}$. Note that (\ref{ac_mse}) is the achievable MSE in the $t$th ADT with given $c_t$, when the BS knows exactly the historical information $\{\phi_n^{(t)}\}_{t = 1}^{t-1}$ and the model parameters $\eta_n$, $\rho_n$. To characterize the feature of (\ref{ac_mse}), we utilize the theorem of variance decomposition, and reformulate (\ref{ac_mse}) as
\begin{equation}
M(c_t) = \mathbb E[{\rm Var}(X^{(t)}|\Phi^{(t)},\Gamma)]-\mathbb E[{\rm Var}(\mathbb E[X^{(t)}|\Phi^{(t)},\Gamma, \{\Phi^{(t)}\}_{t = 1}^{t-1}]|\Phi^{(t)},\Gamma)],\label{mse_rel}
\end{equation}
where the term $\mathbb E[{\rm Var}(X^{(t)}|\Phi^{(t)},\Gamma)]$ is exactly the achievable MSE without the historical knowledge \cite{ChenSparse2018}. To examine the MSE relationship in (\ref{mse_rel}), we expand the second term as
\begin{equation}
\begin{split}
&\mathbb E[{\rm Var}(\mathbb E[X^{(t)}|\Phi^{(t)},\Gamma, \{\Phi^{(t)}\}_{t = 1}^{t-1}]|\Phi^{(t)},\Gamma)] \\
&= \mathbb E_{\Phi^{(t)},\Gamma}[\mathbb E_{\{\Phi^{(t)}\}_{t = 1}^{t-1}}[|\mathbb E[X^{(t)}|\Phi^{(t)},\Gamma, \{\Phi^{(t)}\}_{t = 1}^{t-1}]|^2-|\mathbb E_{\{\Phi^{(t)}\}_{t = 1}^{t-1}}[\mathbb E[X^{(t)}|\Phi^{(t)},\Gamma, \{\Phi^{(t)}\}_{t = 1}^{t-1}]]|^2]].
\end{split}
\end{equation}
Note that we have $\mathbb E[{\rm Var}(\mathbb E[X^{(t)}|\Phi^{(t)},\Gamma, \{\Phi^{(t)}\}_{t = 1}^{t-1}]|\Phi^{(t)},\Gamma)]\ge0$, with equality if,  $$p_{X^{(t)}|\Gamma, \{\Phi^{(t)}\}_{t = 1}^{t-1}}(\cdot) = p_{X^{(t)}|\Gamma}(\cdot).$$ Hence, the achievable MSE of the denoiser can be reduced by $\mathbb E[{\rm Var}(\mathbb E[X^{(t)}|\Phi^{(t)},\Gamma, \{\Phi^{(t)}\}_{t = 1}^{t-1}]|\Phi^{(t)},\Gamma)]$ with given historical knowledge. From the factor graph perspective, this condition holds only in the case that the links between node $h_n^{(t-1)}$, $h_n^{(t)}$ and node $a_n^{(t-1)}$, $a_n^{(t)}$ are vanished. In other words, the achievable MSE with historical knowledge degrades to that without the historical knowledge under same noise level only when there is no temporal correlations between adjacent ADTa. Note that the MSE is a monotone increasing function with respect to the noise level, so that with the inherent temporal correlations, the historical knowledge benefits the detection performance of AMP in each iteration with the same initialization.

To achieve this optimal MSE, one requires to derive the exact posterior distribution or equivalently to track the exact prior distribution $p_{X^{(t)}|\Gamma, \{\Phi^{(t)}\}_{t = 1}^{t-1}}(\cdot)$ in every ADT. However, as we shall see in the next section, such a condition is not impractical, and we turn to find the optimal tractable approximation under some constrains to take advantage of the historical knowledge.

\section{S-AMP: Derivation of Historical Knowledge-Aided Prior}
In this section, we focus on the second part of our schedule, which aims to propagate the beliefs from one ADT to the next, deriving the historical knowledge-aided prior of sparse vector in every ADT. We have claimed that the achievable MSE in each ADT is benefited from the historical knowledge, and the optimal performance is achieved by cycling (\ref{amps})-(\ref{ampe}) until convergence when the prior distribution of sparse signal is in the form of (\ref{lopx}). However, in this section, we will observe that the prior distribution will not maintain a consistent form with ADTs. Worse, the number of mixture components will increase exponentially, making it impractical to accurately track the prior distribution of sparse signal. Thus, in this section, we propose an approximation, restricting the forms of messages in each ADT to make the prior distribution tractable. Then, we find the corresponding approximate historical knowledge-aided prior, which is optimal in the perspective of KL divergence.

We have claimed in section \ref{within} that the generic AMP algorithm in  the ``within'' step of the first part of our schedule decouples the vector-valued estimation problem into a sequence of scalar problems, with the MMSE achieved by tracking the prior distribution $p_{X^{(t)}|\Gamma, \{\Phi^{(t)}\}_{t = 1}^{t-1}}(\cdot)$ in each ADT. This is equivalent to independently tracking the prior distributions $p(x^{(t)}|\eta_n, \rho_n, \{\phi^{(t)}\}_{t = 1}^{t-1})$ for $N$ switching state space models (SSSMs) in each ADT with given historical evidence $\{\phi_n^{(t)}\}_{t = 1}^{t-1}$ and model parameters $\eta_n$, $\rho_n$. Specifically, for each SSSM model related to a particular user device, the evidence $\phi_n^{(t)}$ of Gaussian Markov state-space model (\ref{pht}) is controlled by a Markov chain (\ref{pat}) and the probability measure $p(\phi_n^{(t)}|h_n^{(t)}, a_n^{(t)})= \mathcal{CN}(\phi_n^{(t)}|h_n^{(t)}a_n^{(t)},c_t)$, so that the filter density $p(h_n^{(t)}|\{\phi_n^{(t)}\}_{t = 1}^{t})$
in the $t$th ADT is specified as $p(h_n^{(t)}|\{\phi_n^{(t)}\}_{t = 1}^{t}) = \sum_{a_n^{(t)} = \{0, 1\}}p(a_n^{(t)}|\{\phi_n^{(t)}\}_{t = 1}^{t})p(h_n^{(t)}|a_n^{(t)}, \{\phi_n^{(t)}\}_{t = 1}^{t})$, where
\begin{equation}
p(h_n^{(t)}|a_n^{(t)}, \{\phi_n^{(t)}\}_{t = 1}^{t}) \propto p(\phi_n^{(t)}|h_n^{(t)},a_n^{(t)})\int\nolimits_{h_n^{(t-1)}}
p(h_n^{(t)}|h_n^{(t-1)},a_n^{(t)})p(h_n^{(t-1)}| \{\phi_n^{(t)}\}_{t = 1}^{t-1},a_n^{(t)}),
\end{equation}
\begin{equation}
p(h_n^{(t-1)}| \{\phi_n^{(t)}\}_{t = 1}^{t-1},a_n^{(t)} = k) =\sum\limits_{j = \{0,1\}}\omega_{jk}p(h_n^{(t-1)}|a_n^{(t-1)} = j,
\{\phi_n^{(t)}\}_{t = 1}^{t-1}), \quad k = \{0,1\},
\end{equation}
with the weights given as $\omega_{jk}\propto p_{jk}p(a_n^{(t-1)} = j|\{\phi_n^{(t)}\}_{t = 1}^{t-1})$. Note that we have omitted the conditions $\eta_n$ and $\rho_n$ in the expressions of probability measure, since the parameters $\eta_n$ and $\rho_n$ remain fixed for each user. It is obvious that the component number of the filter density $p(h_n^{(t)}|\{\phi_n^{(t)}\}_{t = 1}^{t})$ will grow exponentially, leading to an intractable form of sparse signal prior distribution after several ADTs.

\subsection{Approximation Strategy}
One approach to solving this problem is to restrict the complexity of the prior distribution representation in every ADT, and find the optimal approximate prior under restrictions, allowing AMP algorithm to operate on it effectively. Specifically, to maintain the sparsity-promoting feature of prior, we choose to restrict the approximate prior in the form of Bernoulli-Gaussian (\ref{lopx}), with the historical knowledge of $\{\phi_n^{(t)}\}_{t = 1}^{t}$ contained in $\hat{\pi}_{n,t}$, $\hat{\xi}_{n,t}$ and
$\hat{\psi}_{n,t}$ in (\ref{lopx}). We notice that the newly derived prior distribution based on the previous approximate prior will typically not in the restricted family, leading that the approximation should be performed in every ADT. One may concern that the errors will be out of control over extended periods of time by accumulation due to the repeated approximations. Fortunately, authors in \cite{Boyen2013} have shown that this problem does not occur because the mere stochasticity of the process serves to attenuate the effects of errors over time, fast enough to prevent the accumulated error from growing unboundedly.

To measure similarity of a distribution and an approximation to it, we introduce the KL-divergence, which is defined as
\begin{equation}
\mathbb D[p(x)||q(x)] = \mathbb E_{p}\left[\ln\frac{p(x)}{q(x)}\right] = \int_xp(x)\ln\frac{p(x)}{q(x)}.\label{kl}
\end{equation}
The integral will be replaced by the summation when the discrete random variable is considered. The KL-divergence, or \emph{Relative Entropy} in the information theory literature \cite{cover2005}, is a very natural measure to quantify the information loss or inefficiency incurred by using distribution $q(x)$ when the true distribution is $p(x)$ \cite{Burnham1998Model}. One important feature of the KL-divergence is that it satisfies $\mathbb D[p(x)||q(x)]\ge 0$, with equality if, and only if, $p(x) = q(x)$, and minimizing (\ref{kl}) can be considered as minimizing the loss of information after approximation.

Hence, one intuition is to minimize the KL-divergence between the newly derived prior distribution and the one restricted in the Gaussian-Bernoulli family in each ADT. However, dealing with the distribution with Gaussian-Bernoulli form in the KL-divergence perspective is not straightforward. The following proposition provide a reasonable alternative to handle this problem.

\emph{Proposition 1}: We define $q(h_n^{(t)},a_n^{(t)}|\{\phi_n^{(t)}\}_{t = 1}^{t})$ as the approximate posterior distribution of user device $n$ in $t$th ADT, and $\tilde p(h_n^{(t)},a_n^{(t)}|\{\phi_n^{(t)}\}_{t = 1}^{t})$ is the derived posterior distribution based on $q(h_n^{(t-1)},a_n^{(t-1)}|\{\phi_n^{(t)}\}_{t = 1}^{t-1})$. Then, the following inequality holds.
\begin{equation}
\begin{split}
\mathbb D[\tilde p(h_n^{(t)},a_n^{(t)}|\{\phi_n^{(t)}\}_{t = 1}^{t})||q(h_n^{(t)},a_n^{(t)}|\{\phi_n^{(t)}\}_{t = 1}^{t})]&\ge \mathbb D[\tilde p(h_n^{(t+1)},a_n^{(t+1)}|\{\phi_n^{(t)}\}_{t = 1}^{t})||q(h_n^{(t+1)},a_n^{(t+1)}|\{\phi_n^{(t)}\}_{t = 1}^{t})]\\
&\ge\mathbb D[\tilde p(x_n^{(t+1)}|\{\phi_n^{(t)}\}_{t = 1}^{t})||q(x_n^{(t+1)}|\{\phi_n^{(t)}\}_{t = 1}^{t})],\label{propo1}
\end{split}
\end{equation}
where we have
\begin{align}
&q(h_n^{(t+1)},a_n^{(t+1)}|\{\phi_n^{(t)}\}_{t = 1}^{t})\triangleq \int_{h_n^{(t)}}\sum_{a_n^{(t)}}q(h_n^{(t)},a_n^{(t)}|\{\phi_n^{(t)}\}_{t = 1}^{t})p(a_n^{(t+1)}|a_n^{(t)})p(h_n^{(t+1)}|h_n^{(t)}),\label{q1}\\
&q(x_n^{(t+1)}|\{\phi_n^{(t)}\}_{t = 1}^{t}) \triangleq \int_{h_n^{(t+1)}}\sum_{a_n^{(t+1)}}q(h_n^{(t+1)},a_n^{(t+1)}|\{\phi_n^{(t)}\}_{t = 1}^{t})\delta(x_n^{(t+1)}-h_n^{(t+1)}a_n^{(t+1)}),\label{q2}\\
&\tilde p(h_n^{(t+1)},a_n^{(t+1)}|\{\phi_n^{(t)}\}_{t = 1}^{t})\triangleq \int_{h_n^{(t)}}\sum_{a_n^{(t)}}\tilde p(h_n^{(t)},a_n^{(t)}|\{\phi_n^{(t)}\}_{t = 1}^{t})p(a_n^{(t+1)}|a_n^{(t)})p(h_n^{(t+1)}|h_n^{(t)}),\\
&\tilde p(x_n^{(t+1)}|\{\phi_n^{(t)}\}_{t = 1}^{t}) \triangleq \int_{h_n^{(t+1)}}\sum_{a_n^{(t+1)}}\tilde p(h_n^{(t+1)},a_n^{(t+1)}|\{\phi_n^{(t)}\}_{t = 1}^{t})\delta(x_n^{(t+1)}-h_n^{(t+1)}a_n^{(t+1)}).
\end{align}
\begin{proof}
Please see the Appendix \ref{app}.
\end{proof}

We can observe that the KL-divergence between the posterior distribution and its approximation never increases by transition through the stochastic processes. This provides us an alternative to dealing with the posterior distribution of $h_n^{(t)}$ and $a_n^{(t)}$ in each ADT instead of the prior of $x_n^{(t)}$. Thus, in one ADT, after the generic AMP framework, we derive the corresponding posterior distribution of $\tilde p(h_n^{(t)},a_n^{(t)}|\{\phi_n^{(t)}\}_{t = 1}^{t})$ and find its optimal approximation from a restricted family in the perspective of KL-divergence.

In this paper, we choose to represent the approximation posterior $q(h_n^{(t)},a_n^{(t)}|\{\phi_n^{(t)}\}_{t = 1}^{t})$ in $t$th ADT using a parametric family that represents as a product of Gaussian and Bernoulli distributions. We will observe that such a representation makes the approximate prior $q(x_n^{(t+1)}|\{\phi_n^{(t)}\}_{t = 1}^{t})$  of $x_n^{(t+1)}$ in $(t+1)$th ADT take the form of Gaussian-Bernoulli (\ref{lopx}). Specifically, the parametric family can be expressed as
\begin{equation}
q(h_n^{(t)},a_n^{(t)}|\{\phi_n^{(t)}\}_{t = 1}^{t}) = q(h_n^{(t)}|\{\phi_n^{(t)}\}_{t = 1}^{t})q(a_n^{(t)}|\{\phi_n^{(t)}\}_{t = 1}^{t}), \label{pa_fa}
\end{equation}
where we have $ q(h_n^{(t)}|\{\phi_n^{(t)}\}_{t = 1}^{t}) = \mathcal{CN}(h_n^{(t)};\bar{\xi}_{n,t},\bar{\psi}_{n,t})$, and $q(a_n^{(t)}|\{\phi_n^{(t)}\}_{t = 1}^{t}) =
\bar{\pi}_{n,t}^{a_n^{(t)}}(1-\bar{\pi}_{n,t})^{1-a_n^{(t)}}$, respectively. Note that the parametric family (\ref{pa_fa}) is an exponential family, so that minimizing the KL-divergence $\mathbb D[\tilde p(h_n^{(t)},a_n^{(t)}|\{\phi_n^{(t)}\}_{t = 1}^{t})||q(h_n^{(t)},a_n^{(t)}|\{\phi_n^{(t)}\}_{t = 1}^{t})]$ is equivalent to adjusting parameters $\bar{\xi}_{n,t}$, $\bar{\psi}_{n,t}$
and $\bar{\pi}_{n,t}$ such that the moments $\mathbb E[h_n^{(t)}]$, $\mathbb E[|h_n^{(t)}|^2]$ and $\mathbb E[a_n^{(t)}]$ match for both distributions $q(h_n^{(t)},a_n^{(t)}|\{\phi_n^{(t)}\}_{t = 1}^{t})$ and $\tilde p(h_n^{(t)},a_n^{(t)}|\{\phi_n^{(t)}\}_{t = 1}^{t})$ \cite{Opper1999}. The moment matching equations can be given as
\begin{align}
\bar{\pi}_{n,t} &= \tilde p(a_n^{(t)} = 1|\{\phi_n^{(t)}\}_{t = 1}^{t}), \label{mo_1} \\
\bar{\xi}_{n,t} &= \int_{h_n^{(t)}}h_n^{(t)}\tilde p(h_n^{(t)}|\{\phi_n^{(t)}\}_{t = 1}^{t}),\label{mo_2}\\
\bar{\psi}_{n,t} &= \int_{h_n^{(t)}}|h_n^{(t)}|^2\tilde p(h_n^{(t)}|\{\phi_n^{(t)}\}_{t = 1}^{t})-|\bar{\xi}_{n,t}|^2,\label{mo_3}
\end{align}
where $\tilde p(h_n^{(t)}|\{\phi_n^{(t)}\}_{t = 1}^{t})$, and $\tilde p(a_n^{(t)} = 1|\{\phi_n^{(t)}\}_{t = 1}^{t})$ are the margins of $\tilde p(h_n^{(t)},a_n^{(t)}|\{\phi_n^{(t)}\}_{t = 1}^{t})$.

\subsection{Derivation of Historical Knowledge-Aided Prior}
We can observe from (\ref{q1}) and (\ref{q2}) that deriving the historical knowledge-aided prior $q(x_n^{(t+1)}|\{\phi_n^{(t)}\}_{t = 1}^{t})$ in $(t+1)$ is equivalent to deriving the approximate posterior distribution $q(h_n^{(t)},a_n^{(t)}|\{\phi_n^{(t)}\}_{t = 1}^{t})$, requiring to specify the corresponding margins in (\ref{mo_1})-(\ref{mo_3}). Recursively, we then derive such margins with the given approximation $q(h_n^{(t-1)},a_n^{(t-1)}|\{\phi_n^{(t)}\}_{t = 1}^{t-1})$ in the previous ADT.

Recalling the factor graph representation of our model in Fig. \ref{graph}, approximating the posterior distribution $\tilde p(h_n^{(t-1)},a_n^{(t-1)}|\{\phi_n^{(t)}\}_{t = 1}^{t})$ by a family $q(h_n^{(t-1)},a_n^{(t-1)}|\{\phi_n^{(t)}\}_{t = 1}^{t-1})$ represented as a product factors is equivalent to disconnecting the link between node $a_n^{(t-1)}$, $h_n^{(t-1)}$. And, restricting the forms of distributions $q(h_n^{(t-1)}|\{\phi_n^{(t)}\}_{t = 1}^{t-1})$ and $q(a_n^{(t-1)}|\{\phi_n^{(t)}\}_{t = 1}^{t-1})$ is equivalent to restricting forms of the messages $\nu_{a_n^{(t-1)}\to q_n^{(t)}}(a_n^{(t-1)})$ and $\nu_{h_n^{(t-1)}\to d_n^{(t)}}(h_n^{(t-1)})$ that pass from $(t-1)$th ADT to the next. Specifically, we have
\begin{align}
&\nu_{a_n^{(t-1)}\to q_n^{(t)}}(a_n^{(t-1)}) = q(a_n^{(t-1)}|\{\phi_n^{(t)}\}_{t = 1}^{t-1}) = \bar{\pi}_{n,t-1}^{a_n^{(t-1)}}(1-\bar{\pi}_{n,t-1})^{1-a_n^{(t-1)}},\label{atq}\\
&\nu_{h_n^{(t-1)}\to d_n^{(t)}}(h_n^{(t-1)}) = q(h_n^{(t-1)}|\{\phi_n^{(t)}\}_{t = 1}^{t-1}) = \mathcal{CN}(h_n^{(t-1)};\bar{\xi}_{n,t-1},\bar{\psi}_{n,t-1}).\label{htd}
\end{align}
According to the structure of Fig. \ref{graph}, the following two messages can be then specified via
\begin{align}
&\nu_{q_n^{(t)}\to a_n^{(t)}}(a_n^{(t)}) = \sum\nolimits_{a_n^{(t-1)}}q_n^{(t)}(a_n^{(t)},a_n^{(t-1)})\nu_{a_n^{(t-1)}\to q_n^{(t)}}(a_n^{(t-1)}),\label{nqta1}\\
&\nu_{d_n^{(t)}\to h_n^{(t)}}(h_n^{(t)}) = \int\nolimits_{h_n^{(t-1)}}d_n^{(t)}(h_n^{(t)},h_n^{(t-1)})
\nu_{h_n^{(t-1)}\to d_n^{(t)}}(h_n^{(t-1)}).\label{ndth1}
\end{align}
We denote $\nu_{q_n^{(t)}\to a_n^{(t)}}(a_n^{(t)} = 1) \triangleq \hat{\pi}_{n,t}$ and $\nu_{d_n^{(t)}\to h_n^{(t)}}(h_n^{(t)}) \triangleq \mathcal{CN}(h_n^{(t)};\hat{\xi}_{n,t},\hat{\psi}_{n,t})$. Plugging (\ref{pat}), (\ref{atq}) into (\ref{nqta1}), and substituting (\ref{pht}), (\ref{htd}) into (\ref{ndth1}), we obtain
\begin{equation}
\begin{split}
&\hat{\pi}_{n,t}= p_{10}(1-\bar{\pi}_{n,t-1})+(1-p_{01})\bar{\pi}_{n,t-1},\\
&\hat{\xi}_{n,t}=\eta_n\bar{\xi}_{n,t-1},\quad\hat{\psi}_{n,t}=\eta_n^2\bar{\psi}_{n,t-1}+(1-\eta_n^2)\rho_n.\label{pa_pr}
\end{split}
\end{equation}

We emphasize again that we focus on a filtering-like problem, where only the knowledge of the previous ADTS can be utilized in current ADT, and we can omit the links between nodes $a_n^{(t)}$, $a_n^{(t+1)}$ as well as $h_n^{(t)}$, $h_n^{(t+1)}$ for each $n$ when performing inference in the $t$th ADT. Then the corresponding margins in (\ref{mo_1})-(\ref{mo_3}) can be specified via
\begin{align}
&\tilde p(a_n^{(t)}|\{\phi_n^{(t)}\}_{t = 1}^{t}) \propto \nu_{f_n^{(t)}\to a_n^{(t)}}(a_n^{(t)})\nu_{q_n^{(t)}\to
a_n^{(t)}}(a_n^{(t)}),\label{fla1}\\
&\tilde p(h_n^{(t)}|\{\phi_n^{(t)}\}_{t = 1}^{t}) \propto \nu_{f_n^{(t)}\to h_n^{(t)}}(h_n^{(t)})\nu_{d_n^{(t)}\to
h_n^{(t)}}(h_n^{(t)}).\label{flh1}
\end{align}
Specifically, the explicit expressions of the following messages can be obtained as
\begin{align}
&\nu_{f_n^{(t)}\to a_n^{(t)}}(a_n^{(t)}) \propto \int_{h_n^{(t)}}\int_{x_n^{(t)}}f_n^{(t)}(x_n^{(t)}, a_n^{(t)},
h_n^{(t)})\cdot\nu_{x_n^{(t)}\to f_n^{(t)}}(x_n^{(t)})\cdot\nu_{h_n^{(t)}\to f_n^{(t)}}(h_n^{(t)})\label{fta1},\\
&\nu_{f_n^{(t)}\to h_n^{(t)}}(h_n^{(t)})\propto\sum\limits_{a_n^{(t)} = \{0,1\}}\int_{x_n^{(t)}}f_n^{(t)}(x_n^{(t)}, a_n^{(t)},
h_n^{(t)})\cdot\nu_{x_n^{(t)}\to f_n^{(t)}}(x_n^{(t)})\cdot\nu_{a_n^{(t)}\to f_n^{(t)}}(a_n^{(t)}).
\end{align}
Under the structure of our factor graph, it is obvious that the following two relationships always hold, e.g., $\nu_{a_n^{(t)}\to f_n^{(t)}}(a_n^{(t)}) = \nu_{q_n^{(t)}\to a_n^{(t)}}(a_n^{(t)})$ and $\nu_{h_n^{(t)}\to f_n^{(t)}}(h_n^{(t)}) = \nu_{d_n^{(t)}\to h_n^{(t)}}(h_n^{(t)})$. As a consequence, we specify the posterior distributions (\ref{fla1}) and (\ref{flh1}) as
\begin{align}
&\tilde p(a_n^{(t)}|\{\phi_n^{(t)}\}_{t = 1}^{t}) \propto
(\hat{\pi}_{n,t}\tilde{\pi}_{n,t})^{a_n^{(t)}}[(1-\hat{\pi}_{n,t})(1-\tilde{\pi}_{n,t})]^{1-a_n^{(t)}},\label{posa}\\ &\tilde p(h_n^{(t)}|\{\phi_n^{(t)}\}_{t = 1}^{t}) \propto
\hat{\pi}_{n,t}\tilde{\pi}_{n,t}\mathcal{CN}(h_n^{(t)};\tilde{\tau}_{n,t},\tilde{\kappa}_{n,t})+(1-\hat{\pi}_{n,t})(1-\tilde{\pi}_{n,t})
\mathcal{CN}(h_n^{(t)};\hat{\xi}_{n,t}, \hat{\psi}_{n,t}),\label{posh}
\end{align}
where we define
\begin{align}
&\tilde{\pi}_{n,t}\triangleq\left(1+\left(\frac{\hat{\pi}_{n,t}}{1-\hat{\pi}_{n,t}}\right)\gamma_{n,t}(\phi_n^{(t)},c_t)\right)^{-1},\label{fta2}\\
&\tilde{\kappa}_{n,t}\triangleq
\frac{c_t\hat{\psi}_{n,t}}{c_t+\hat{\psi}_{n,t}},\quad\tilde{\tau}_{n,t}\triangleq\tilde{\kappa}_{n,t}\cdot\left(\frac{\phi_n^{(t)}}{c_t}+\frac{\hat{\xi}_{n,t}}{\hat{\psi}_{n,t}}\right)\label{tau}.
\end{align}

After that, we can obtain the explicit expression of the approximation distribution $q(h_n^{(t)},a_n^{(t)}|\{\phi_n^{(t)}\}_{t = 1}^{t})$ represented as the form (\ref{pa_fa}) by combining the moment matching equations (\ref{mo_1})-(\ref{mo_3}), and the marginal expressions (\ref{posa}), (\ref{posh}). As a consequence, we have
\begin{align}
&\bar{\pi}_{n,t} = \frac{\hat{\pi}_{n,t}\tilde{\pi}_{n,t}}{\hat{\pi}_{n,t}\tilde{\pi}_{n,t}+(1-\hat{\pi}_{n,t})(1-\tilde{\pi}_{n,t})},\quad
\bar{\xi}_{n,t} = \bar{\pi}_{n,t}\tilde{\tau}_{n,t}+(1-\bar{\pi}_{n,t})\hat{\xi}_{n,t},\\
&\bar{\psi}_{n,t} =
\bar{\pi}_{n,t}(|\tilde{\tau}_{n,t}|^2+\tilde{\kappa}_{n,t})+(1-\bar{\pi}_{n,t})(|\hat{\xi}_{n,t}|^2+\hat{\psi}_{n,t})-|\bar{\xi}_{n,t}|^2.
\end{align}

Then, the historical knowledge aided-prior can be formulated via (\ref{q1}) and (\ref{q2}). According to (\ref{atq}) and (\ref{htd}), executing (\ref{q1}) is equivalent to implementing (\ref{nqta1}), (\ref{ndth1}) and further implementing
\begin{equation}
\begin{split}
 q(h_n^{(t+1)},a_n^{(t+1)}|\{\phi_n^{(t)}\}_{t = 1}^{t}) = &q(a_n^{(t+1)}|\{\phi_n^{(t)}\}_{t = 1}^{t})q(h_n^{(t+1)}|\{\phi_n^{(t)}\}_{t = 1}^{t}) \\
 = &\nu_{q_n^{(t+1)}\to
a_n^{(t+1)}}(a_n^{(t+1)})\nu_{d_n^{(t+1)}\to
h_n^{(t+1)}}(h_n^{(t+1)}).
\end{split}
\end{equation}
In addition, we notice that the equation (\ref{q2}) is equivalent to (\ref{lo_prx}) with index $(t+1)$. Observing (\ref{nqta1}) and (\ref{ndth1}), the messages $\nu_{q_n^{(t)}\to a_n^{(t)}}(a_n^{(t)} = 1)$ and $\nu_{d_n^{(t)}\to h_n^{(t)}}(h_n^{(t)})$ are in the same forms as (\ref{dth}) assumed in section \ref{into}. Hence, the historical knowledge aided-prior of the sparse vector $\boldsymbol x^{(t)}$ in each $t$th ADT will maintain the consistent form of Gaussian-Bernoulli (\ref{lopx}) with the corresponding parameters specified by (\ref{pa_pr}). This leads that the generic AMP update equations in each ADT are (\ref{amps})-(\ref{ampe}). Somewhat differently, since the equation (\ref{ampe}) is equivalent to the state evolution equation (\ref{ge_st}) only when the exact prior distribution of $\boldsymbol x^{(t)}$ is in the form of (\ref{lopx}), the approximations in our proposed method will cause the loss of accuracy. To be more accurate, we consider an empirical alternative of (\ref{ampe}), given as $c^{i+1}=\frac{1}{\sqrt{L}}||\boldsymbol z^{i+1}||_2$, where $\boldsymbol z^{i+1}\triangleq [z^{i+1}_1, z^{i+1}_2,\dots, z^{i+1}_L]^{T}\in\mathbb C^{L}$, which can be used as an efficient approximation of the state evolution \cite{Montanari2011}.
\begin{algorithm}[t]
\caption{Proposed S-AMP algorithm}
\hspace*{0.02in}{\bf Input:}
Received Signal: $\{\boldsymbol y^{(t)}\}_{t = 1}^{T}$;\\
Pilot matrix: $\boldsymbol S$.\\
\hspace*{0.02in}{\bf Output:}
Recovered sparse vector $\{\boldsymbol {\hat x}^{(t)}\}_{t = 1}^{T}$.
\begin{algorithmic}[1]
\For{$t\le T$}
\State Initialization: $\mu_{n}^{0} = 0$, $z_{l}^{0} = y_l$, and $c^{0}\gg\sigma_w^2$ for all $n$ and $l$ (we drop the index of $t$).
\For{$i\le I$}
\State $\forall n$, calculate AMP equations with $\hat{\pi}_{n,t}$, $\hat{\xi}_{n,t}$ and $\hat{\psi}_{n,t}$.
$$\phi_n^{i} = \sum\nolimits_{l = 1}^{L}S_{ln}^*z_{l}^i+\mu_{n}^{i},\quad\mu_{n}^{i+1} = F_n(\phi_n^{i},c^{i}),\quad v_{n}^{i+1} = G_n(\phi_n^{i},c^{i}),$$
$$z_{l}^{i+1}= y_l-\sum\nolimits_{n = 1}^{N}S_{ln}\mu_{n}^{i+1}+\frac{z_{l}^{i}}{L}\sum\nolimits_{n = 1}^NF_n'(\phi_n^{i},c^{i}),\quad c^{i+1}=\frac{1}{\sqrt{L}}||\boldsymbol z^{i+1}||_2,$$
\State
\EndFor
\State\Return $\boldsymbol {\hat x}^{(t)} = \boldsymbol\mu^I$.
\State $\forall n$, calculate the following parameters:
$$\tilde{\pi}_{n,t}\triangleq\left(1+\left(\frac{\hat{\pi}_{n,t}}{1-\hat{\pi}_{n,t}}\right)
\gamma_{n,t}(\phi_n^{(t)},c_t)\right)^{-1},$$
$$\tilde{\kappa}_{n,t}\triangleq\frac{c_t\hat{\psi}_{n,t}}{c_t+\hat{\psi}_{n,t}},
\quad\tilde{\tau}_{n,t}\triangleq\tilde{\kappa}_{n,t}\cdot\left(\frac{\phi_n^{(t)}}{c_t}+\frac{\hat{\xi}_{n,t}}{\hat{\psi}_{n,t}}\right),$$
$$\bar{\pi}_{n,t} = \frac{\hat{\pi}_{n,t}\tilde{\pi}_{n,t}}{\hat{\pi}_{n,t}\tilde{\pi}_{n,t}+(1-\hat{\pi}_{n,t})(1-\tilde{\pi}_{n,t})},\quad
\bar{\xi}_{n,t} = \bar{\pi}_{n,t}\tilde{\tau}_{n,t}+(1-\bar{\pi}_{n,t})\hat{\xi}_{n,t},$$
$$\bar{\psi}_{n,t} =
\bar{\pi}_{n,t}(|\tilde{\tau}_{n,t}|^2+\tilde{\kappa}_{n,t})+(1-\bar{\pi}_{n,t})(|\hat{\xi}_{n,t}|^2+\hat{\psi}_{n,t})-|\bar{\xi}_{n,t}|^2,$$
$$\hat{\pi}_{n,t+1}= p_{10}(1-\bar{\pi}_{n,t})+(1-p_{01})\bar{\pi}_{n,t},$$
$$\hat{\xi}_{n,t+1}=\eta_n\bar{\xi}_{n,t},\quad\hat{\psi}_{n,t+1}=\eta_n^2\bar{\psi}_{n,t}+(1-\eta_n^2)\rho_n.$$
\EndFor

\end{algorithmic}
\end{algorithm}

We summary our proposed S-AMP in Algorithm 1. The main computational burden of the proposed method is the generic AMP equations in section \ref{within}, resulting the complexity in each iteration as $\mathcal O(MN)$, which is equivalent to that of the algorithms proposed in \cite{LiuMassive12018, ChenSparse2018}.

So far, we have given all the derivations of our proposed S-AMP algorithm. In the following, we will consider the concrete active user detection and channel estimation strategy. As we shall see, based on our proposed S-AMP algorithm, the user detector and channel estimator can be designed directly.

\section{Active User Detection and Channel Estimation}
After the generic AMP equations in the proposed S-AMP algorithm converges in each ADT, we focus on the joint active user detection and channel estimation. The hypothesis testing to find out the active device is given by
\begin{equation}
\begin{cases}
H_1:\quad  & a_n^{(t)} = 1, \mbox{ active device}; \\
H_0:\quad  & a_n^{(t)} = 0, \mbox{ inactive device}.
\end{cases}
\end{equation}
Applying (\ref{ndth1}), the LLR test rule given decision threshold $l_{n,t}$ in the $t$th ADT is given by
\begin{equation}
{\rm LLR} = \log\left(\frac{\tilde p(\phi_n^{(t)}|a_n^{(t)} = 1,\{\phi_n^{(t)}\}_{t =
1}^{t-1})}{\tilde p(\phi_n^{(t)}|a_n^{(t)} = 0,\{\phi_n^{(t)}\}_{t = 1}^{t-1})} \right)\mathop{\gtrless}\limits_{H_0}^{H_1} l_{n,t},\label{LLR}
\end{equation}
where we define the approximate likelihood function as $$\tilde p(\phi_n^{(t)}|a_n^{(t)}, \{\phi_n^{(t)}\}_{t = 1}^{t-1})\triangleq \int_{h_n^{(t)}}p(\phi_n^{(t)}|h_n^{(t)}, a_n^{(t)})\nu_{d_n^{(t)}\to h_n^{(t)}}(h_n^{(t)}).$$ Specifically, we have
\begin{align}
&\tilde p(\phi_n^{(t)}|a_n^{(t)} = 0, \{\phi_n^{(t)}\}_{t = 1}^{t-1}) = \mathcal{CN}(\phi_n^{(t)};0,c_t)\label{h0},\\
&\tilde p(\phi_n^{(t)}|a_n^{(t)} = 1, \{\phi_n^{(t)}\}_{t = 1}^{t-1}) = \mathcal{CN}(\phi_n^{(t)};\hat{\xi}_{n,t},\hat{\psi}_{n,t}+c_t).\label{h1}
\end{align}
We notice that this is a general Gaussian hypothesis testing problem \cite{TreesDetection2001}. As a consequence, the sufficient statistic for the LLR detector in the $t$th ADT for each user $n$ is given as
\begin{equation}
\mathcal T(\phi_n^{(t)}) = |\phi_n^{(t)}|^2+2c_t\hat{\psi}_{n,t}^{-1}\mathfrak R(\hat{\xi}_{n,t}^*\phi_n^{(t)})+c_t^2\hat{\psi}_{n,t}^{-2}|\hat{\xi}_{n,t}|^2 =
|\phi_n^{(t)}+c_t\hat{\psi}_{n,t}^{-1}\hat{\xi}_{n,t}|^2,\label{quat}
\end{equation}
which contains both the quadratic and linear term of variable $\phi_n^{(t)}$. To examine the functional form of the sufficient statistic $\mathcal T(\phi_n^{(t)})$, we see that if we assume the zero mean of the local prior, i.e., $\hat{\xi}_{n,t} = 0$, the proposed detector transforms to the traditional energy detector derived in \cite{LiuMassive12018, ChenSparse2018}.

Another crucial issue for the detector is the design of decision criterion. In the detection literature, the basic criteria are
the Neyman-pearson and the Bayes criterion \cite{TreesDetection2001}. In this paper, we consider the Bayes criterion for the following reasons. First, the Bayes criterion focuses on the prior knowledge, which is neglected by the Neyman-pearson criterion. Second, we mainly concerned the detection error probability in our scenario, which is defined as the sum of false alarm and missed detection probabilities, and utilizing the Bayes criterion will achieve minimum detection error probability when given the exact prior distribution. Further, we can observe that the sufficient statistic $\mathcal T(\phi_n^{(t)})$ in (\ref{quat}) is derived based on the parameters $\hat{\xi}_{n,t}$ ,$\hat{\psi}_{n,t}$ ,$c_t$, which include the historical knowledge of $\{\phi_n^{(t)}\}_{t = 1}^{t-1}$ and cannot be predicted beforehand. Since the detection threshold under Neyman-Pearson criterion is designed based on the map between performance metrics, which is not explicit because of the linear term in $\mathcal T(\phi_{nt})$ \cite{TreesDetection2001}, it is impractical to instantaneously design the threshold.  On the contrary, under the Bayes criterion, the LLR test is equivalent to computing the posterior probabilities of two hypotheses and choose the large one. Note that under S-AMP framework, the acquisition of the approximation posterior probabilities is very straightforward, as we have derived in (\ref{posa}). Specifically, the detector under Bayes criterion can be written as
\begin{equation}
\begin{cases}
H_1, \quad &\mbox{if}\quad \frac{\tilde p(a_n^{(t)} = 1|\{\phi_n^{(t)}\}_{t = 1}^{t})}{\tilde p(a_n^{(t)} = 0|\{\phi_n^{(t)}\}_{t =
1}^{t})}\ge1,\\
H_0, \quad &\mbox{otherwise}.
\end{cases}
\end{equation}

After active user detection, the channel estimation for the active users can be implemented directly by applying the recovered sparse vector as the estimated channel, e.g., we consider $\hat h_n^{(t)} = \hat x_n^{(t)}$ for each user device $n$, where $\hat x_n$ is the corresponding element of $\boldsymbol {\hat x}^{(t)}$.

\begin{figure}[!t]
\centering
\includegraphics[width=4in]{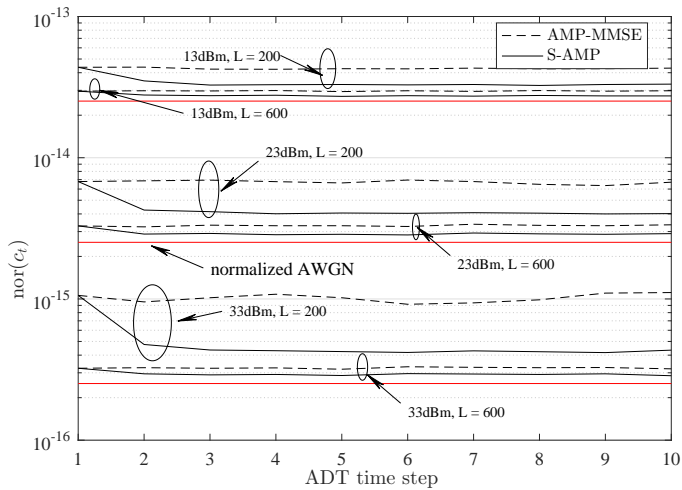}
\caption{Normalized fix point of state evolution function.}
\label{ct}
\end{figure}

\section{Numerical Results}
In this section, some numerical examples are provided to verify our theoretical results. We simulate the mMTC system with $N = 2000$ devices. The user devices are randomly located in a cell with distance $d_n$ for $n$th user. It is assumed that $d_n$, $n = 1,\dots,N$ are randomly distributed in the region $[0.05$km, $1$km$]$. The large-scale fading between each user and the BS is considered as $\beta_n = -128.1-36.7\log_{10}(d_n)$ in dB. The
power spectral density of the AWGN at the BS is set to be $-169$dBm/Hz with the wireless channel bandwidth $10$MHz. The duration of ADP is denoted as $T_s$, and the total number of ADPs is set as $20$. For simplification, we
consider the same transmission power for each user, i.e., $P_n = P$. The average speed for each user is supposed randomly distributed in $[0,50]$km/h, the carrier frequency is set as $3.5$GHz. Since the temporal correlation of user indicator is specified by $\lambda$ and
the transition probability $p_{01}$, we assume $p_{01} = r(1-\lambda)$ and $p_{10} = r\lambda$, where $r$ denote a scale factor that controls the specific value of transition probability. For the performance metrics, we utilize the normalized mean squared error (NMSE) to illustrate the sparse vector recovery and channel estimation performance. And, the active user detection performance is measured by detection error probability (DEP). We first set the access probability of users in each ADT is assumed to be $\lambda = 0.05$, the scale $r = 0.1$ and the duration of ADP is $T_s = 100$us.

To show the advantage of the proposed S-AMP algorithms, we compare our results with the conventional CS-based algorithm. In the non-Bayesian framework, we consider the classical OMP \cite{OMP2007} algorithm, and the R-PIA-ASP \cite{Efficient2017}, which is the most efficient existing method utilizes temporal correlation of the user support. Further, we consider the oracle LS algorithm, which assumes the true active user support set is exactly known at the BS. In the Bayesian literature, we consider the classical AMP algorithm \cite{DonohoMessage2010} and the AMP-MMSE algorithm \cite{LiuMassive12018, ChenSparse2018} as the counterparts of our proposed S-AMP.

\begin{figure}[!t]
\centering
\includegraphics[width=4in]{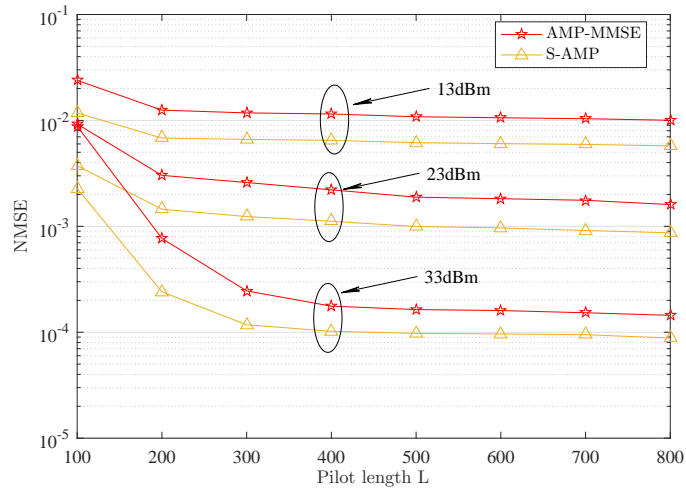}
\caption{NMSE performance of sparse vector recovery.}
\label{xvl}
\end{figure}

Fig. {\ref{ct}} depicts the normalized fix point of state evolution equation from $t = 1$ to $t = 10$ of both S-AMP and AMP-MMSE algorithms under different settings. The normalized fix point of state evolution equation in $t$the ADT is defined as ${\rm nor}(c_t)\triangleq \frac{P}{P_0}c_t$, where $P_0 = 13$dBm is the reference level of power. We can observe that in the first ADT, where each of two algorithms has the historical knowledge, ${\rm nor}(c_t)$ of the two algorithms are almost equal. On the contrary, with given historical knowledge, ${\rm nor}(c_t)$ of S-AMP is lower than that of AMP-MMSE. In addition, under low power level, the performance gain provided by pilot length shrinks compared with that under high power level, since the AMGN term in state evolution function is dominant in this case. Further, with the pilot length increases, ${\rm nor}(c_t)$ of both the two algorithms will converge to the normalized AWGN level, defined as $\frac{P}{P_0}\sigma_w^2$, which is the lower bound of state evolution fix point. Hence, as $L$ increases, the performance gain of the S-AMP compared with the AMP-MMSE will come more from the prior distribution.
\begin{figure*}

\subfigure[Detection performance]
{
\begin{minipage}[b]{0.5\textwidth}
\includegraphics[width=3.2 in]{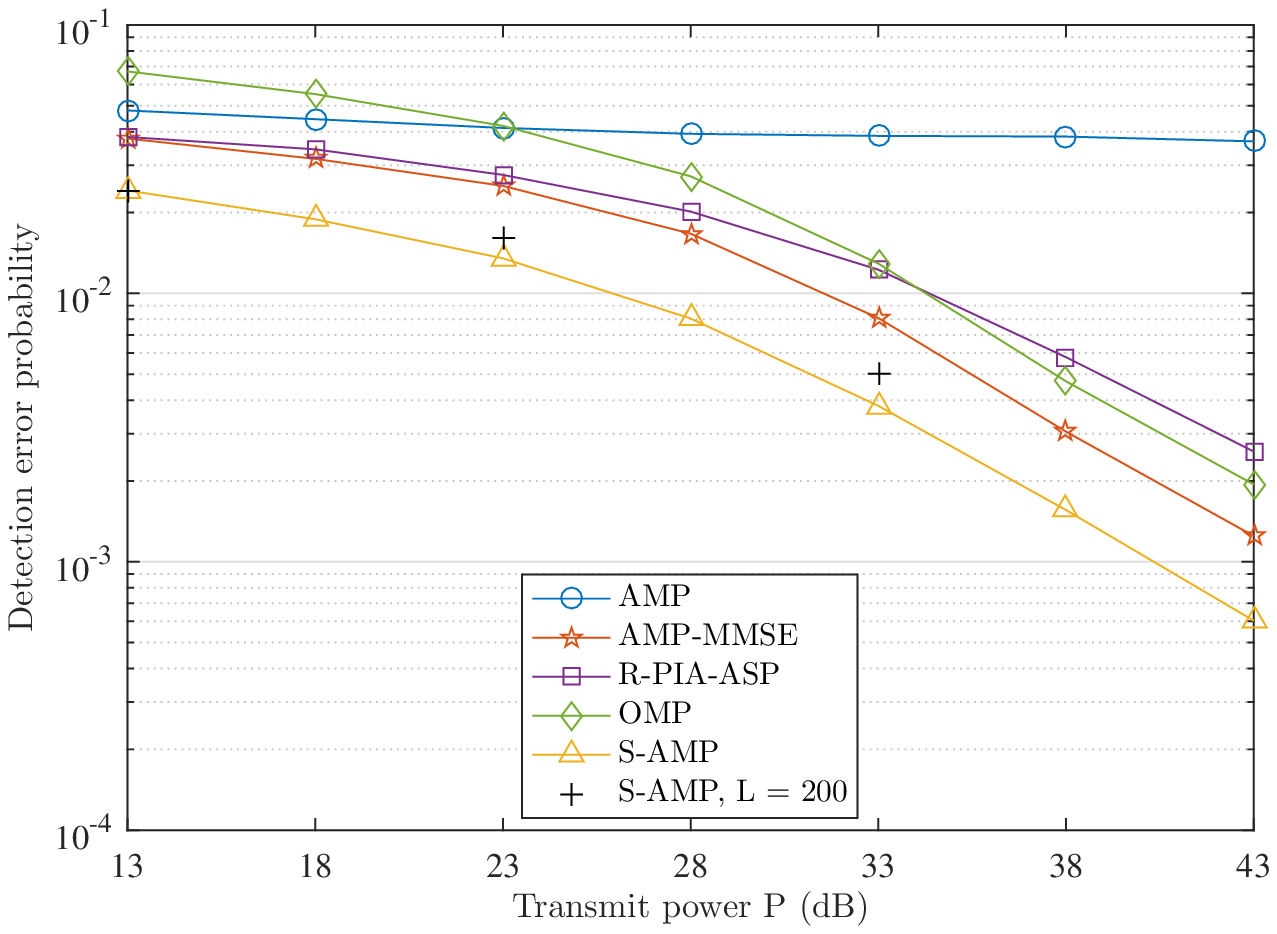}
\end{minipage}
}
\subfigure[Channel estimation performance]
{
\begin{minipage}[b]{0.5\textwidth}
\includegraphics[width=3.2 in]{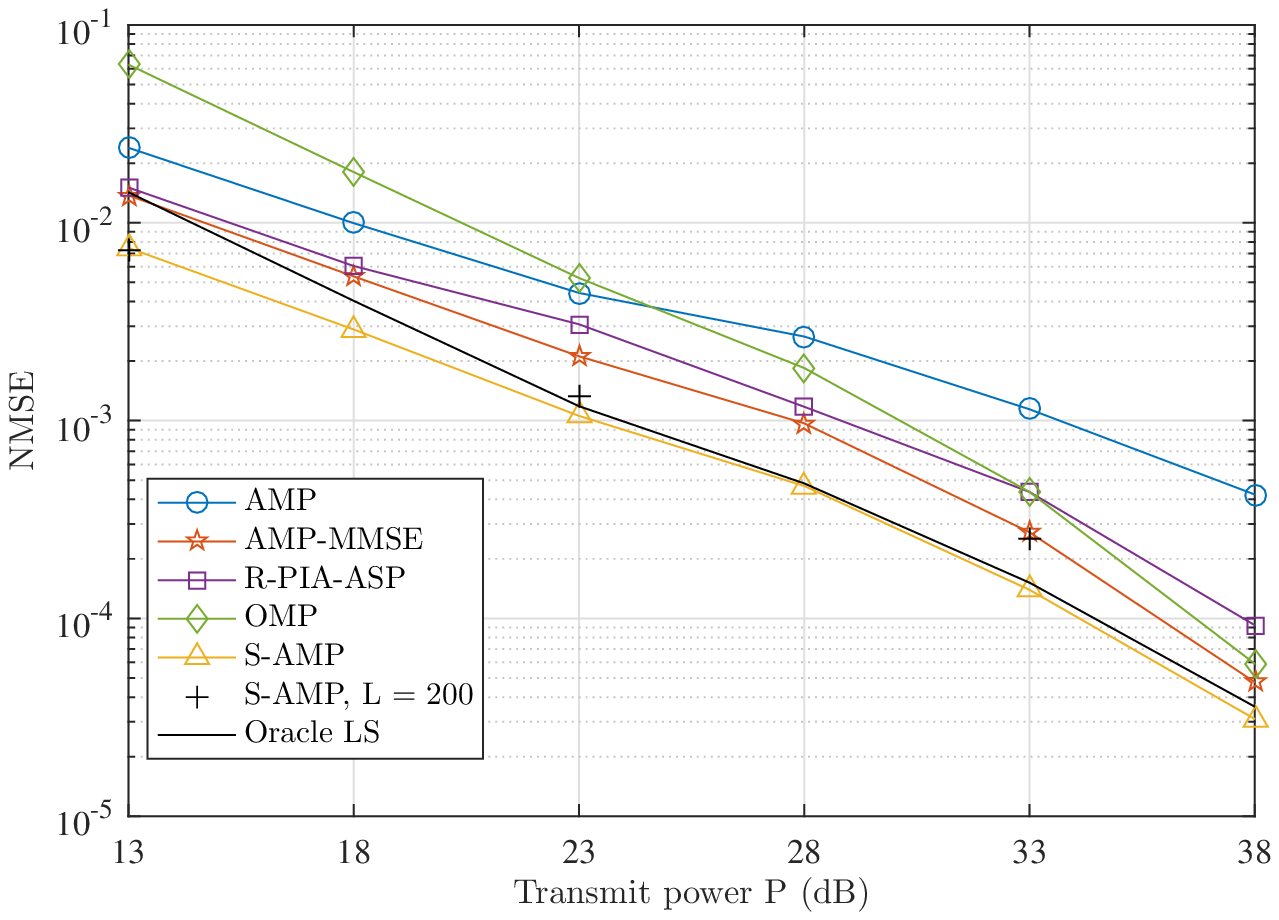}
\end{minipage}
}
\caption{Performance comparisons with respect to the transmission power.}
\label{hvp}
\end{figure*}

Fig. {\ref{xvl}} depicts effect of pilot length and power level on sparse vector recovery performance of the AMP-MMSE and S-AMP algorithms. We can find that the S-AMP shows lower NMSE than AMP-MMSE for all settings, verifying that the historical knowledge benefits the AMP framework. We notice that in all the power levels, when length of pilot is relative large, increasing $L$ will not distinctly increase the NMSE performance. To explain, we compare the Fig. \ref{xvl} with Fig. {\ref{ct}}, and notice that with the pilot length increases, ${\rm nor}(c_t)$ in Fig. {\ref{ct}} of both the two algorithms will converge to the normalized AWGN level, defined as $\frac{P}{P_0}\sigma_w^2$, which is the lower bound of state evolution fix point. Hence, as $L$ increases, the performance gain of the S-AMP compared with the AMP-MMSE will come more from the prior distribution. We turn to the Fig. {\ref{xvl}} and we can observe that in the region of higher number of pilot, the S-AMP algorithm achieves about $3$dB NMSE performance gain compared with its counterpart. This result indicates the clear advantage of prior with the aid of historical knowledge.

\begin{figure*}

\subfigure[Detection performance]
{
\begin{minipage}[b]{0.48\textwidth}

\includegraphics[width=3.2 in]{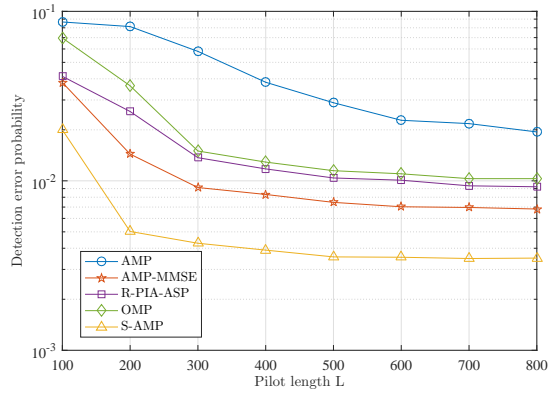}
\end{minipage}
}
\subfigure[Channel estimation performance]
{
\begin{minipage}[b]{0.48\textwidth}

\includegraphics[width=3.2 in]{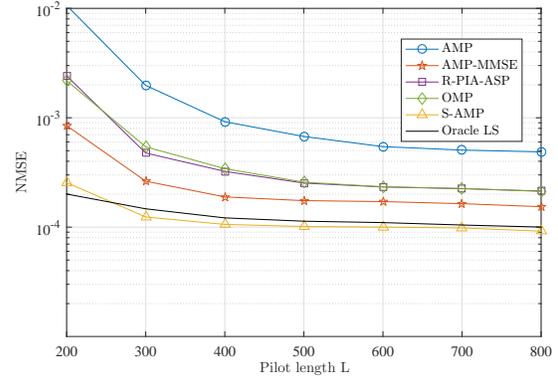}
%
\end{minipage}
}
\caption{Performance comparisons with respect to the pilot length.}
\label{hvl}
\end{figure*}

\begin{figure*}
\centering
\subfigure[ADP duration: 10ms]{
\begin{minipage}[b]{0.48\textwidth}
\includegraphics[width=3.2 in]{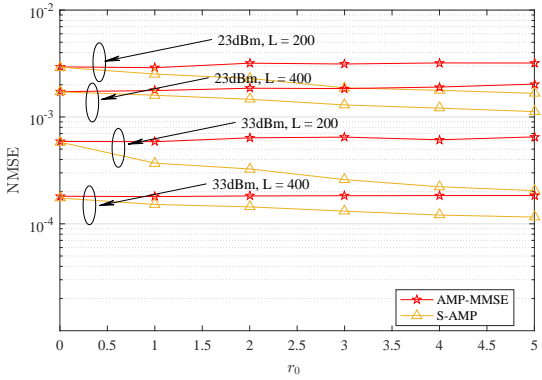}\\
\includegraphics[width=3.2 in]{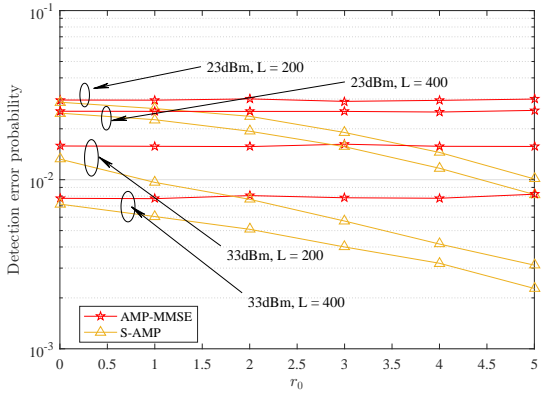}
\end{minipage}}
\subfigure[ADP duration: 100us]{
\begin{minipage}[b]{0.48\textwidth}
\includegraphics[width=3.2 in]{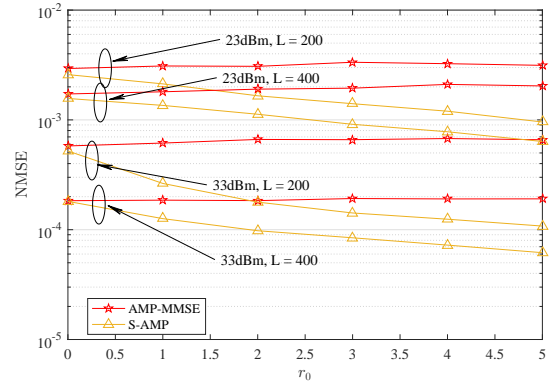}\\
\includegraphics[width=3.2 in]{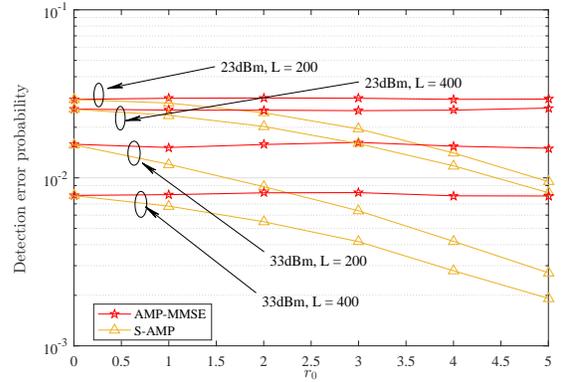}
\end{minipage}}
\caption{Performance of S-AMP versus user state transition probability in different cases of ADP durations. Top: channel estimation performance; Bottom: active user detection performance.}
\label{r_adt}
\end{figure*}
Next, we investigate the active user detection and channel estimation performance of the S-AMP. Fig. {\ref{hvp}} provides the user detection and channel estimation performances comparisons of the considered baseline algorithms versus the power level of the user devices. The pilot length is set as $L = 400$. We adopt the detector in \cite{HannakJoint2015} for AMP algorithm, and the detections for OMP and R-PIA-ASP are based on their estimated support. For AMP-MMSE and S-AMP, we consider the Bayesian detector. We see obviously that our proposed S-AMP outperforms its counterparts in the both detection and channel estimation performances under all considered settings. With prior information, the Bayesian CS algorithms, such as AMP-MMSE, S-AMP obtain distinct performances gain compared with the non-Bayesian methods.  And, with historical knowledge-aided prior, our S-AMP algorithm further improves the performance compared with the AMP-MMSE algorithm. Specifically, for detection, the DEP is reduced by half, and for channel estimation, about $3$dB NMSE gain is achieved. Particularly, the S-AMP even outperforms the Oracle LS method in channel estimation, which provides the lower bound of any non-Bayesian method.

Fig. \ref{hvl} investigate the detection and channel estimation performance with respect to the pilot length. The transmission power is set as $33$dBm. We can find that the user detection and channel estimation performances are improved with the increase of the pilot length. In the region of lower pilot, the DEP and NMSE reduce very fast, leading that the proposed S-AMP algorithm achieves significant performance even with low pilot overhead. For channel estimation, there is a slight performance loss of S-AMP compared with the Oracle LS, when $L = 200$, since the Oracle LS knows exactly the support set. However, in all region, the proposed S-AMP algorithm achieves lower NMSE than any other its achievable counterparts.

Fig. \ref{r_adt} demonstrates the performance of S-AMP under different settings of ADP durations versus $r_0$, which is specified by $r = 1/2^{r_0}$, controlling the transition probability of the user state. We note that the probability of the user switching state will decrease as $r_0$ increases. We can observe from Fig. \ref{r_adt} that with $r_0$ increases, both the channel estimation error and detection error of the S-AMP decrease. On the contrary, the variation of $r_0$ has no impact on AMP-MMSE, since it does not utilize the temporal structure. On the other hand, we note that the duration of ADP controls the temporal correlation of user channel, and a large duration associates with a poor temporal correlation of user channel. We can see from Fig. \ref{r_adt} that considering a small ADP duration will enhance the channel estimation performance while having few impacts on the detection performance.

\begin{figure*}
\subfigure[Channel estimation performance]{
\begin{minipage}[b]{0.48\textwidth}
\includegraphics[width=3.2 in]{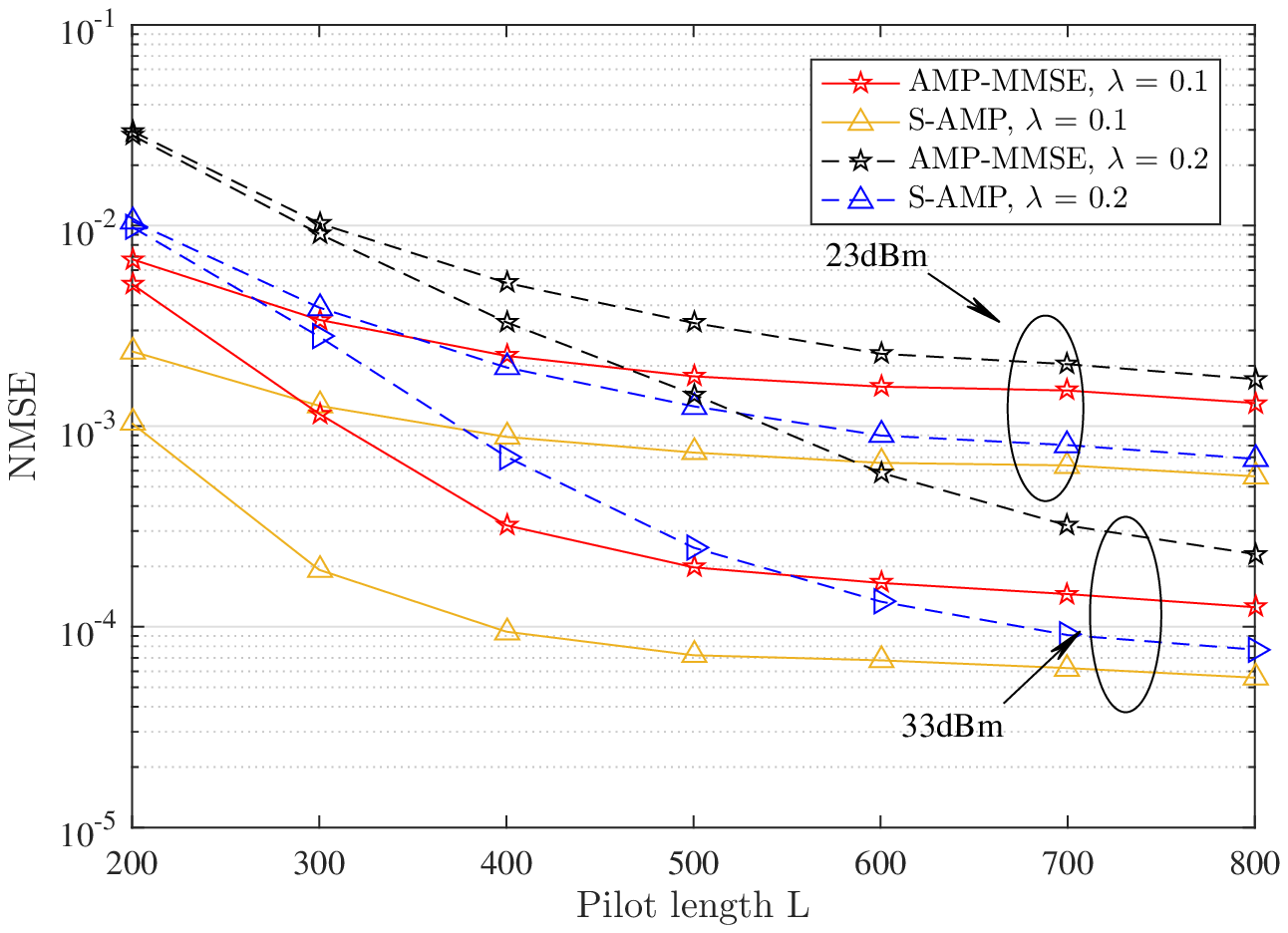}
\end{minipage}
}
\subfigure[Detection performance]{
\begin{minipage}[b]{0.48\textwidth}
\includegraphics[width=3.2 in]{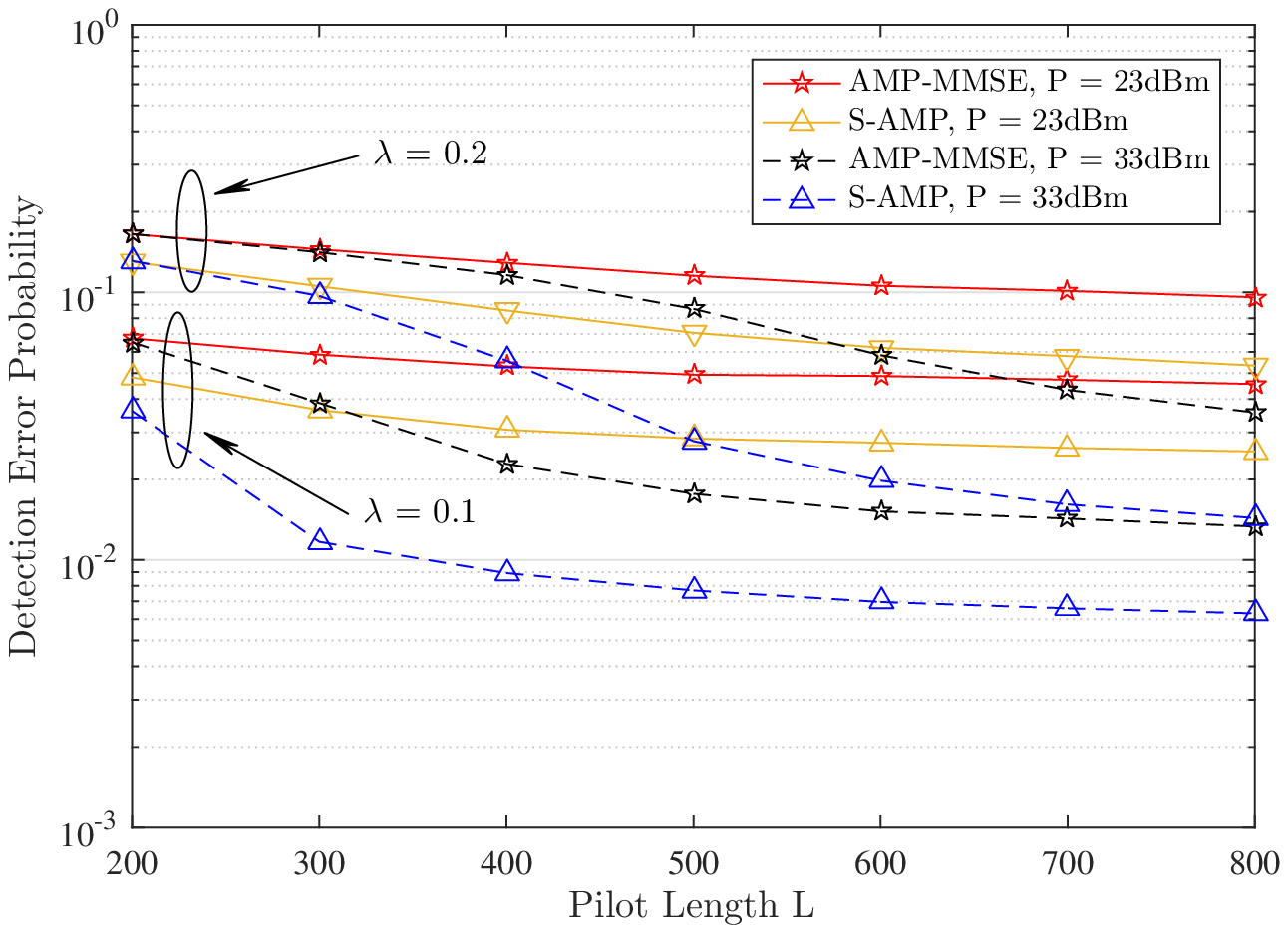}
\end{minipage}
}
\caption{Performance of S-AMP in different cases of access probabilities.}
\label{lam_mse}
\end{figure*}

Fig. \ref{lam_mse} investigate the channel estimation and detection performance of S-AMP under different settings of access probabilities $\lambda$. We can see that in the same settings of pilot length and transmit power, increasing the sparsity level $\lambda$ will decrease both the performances of channel estimation and detection. However, our proposed S-AMP algorithm outperforms the AMP-MMSE in all settings, indicating the scalability of the S-AMP in different sparsity levels. In addition, we can intuitively see that the impacts of improving sparsity level will be compensated by increasing the length of pilot.

\section{Conclusion}
In this paper, we proposed a grant-free novel MRA strategy, which considered both the sporadic traffic and short packet features of mMTC scenario. Such a strategy results in the temporal correlation of the access state spare vector, leading to that the joint user detection and channel estimation formed a dynamic CS problem. We therefore proposed a novel S-AMP algorithm to sequentially recover the spare vector. Further, we derived the Bayes detector for active user detection and corresponding channel estimator based on the S-AMP. We verified that the S-AMP outperforms the traditional AMP algorithms and other non-Bayes methods in both active user detection and channel estimation performances under our scenario, indicating the clear advantage of accounting for temporal correlation of the access state spare vector in user activity detector and channel estimator design.

\appendix
\subsection{Proof of Proposition 1}\label{app}
Before proceeding, we define $\chi_n^{(t)}\triangleq\{h_n^{(t)},a_n^{(t)}\}$ and $\boldsymbol \phi_n^{(t)}\triangleq\{\phi_n^{(t)}\}_{t = 1}^{t}$, utilizing the definition of KL-divergence, we have
\begin{equation}
\begin{split}
& \mathbb D[\tilde p(\chi_n^{(t)}|\boldsymbol \phi_n^{(t)})||q(\chi_n^{(t)}|\boldsymbol \phi_n^{(t)})]\\
= &\int_{\chi_n^{(t)}}\tilde p(\chi_n^{(t)}|\boldsymbol \phi_n^{(t)})\log \frac{\tilde p(\chi_n^{(t)}|\boldsymbol \phi_n^{(t)})}{q(\chi_n^{(t)}|\boldsymbol \phi_n^{(t)})}\\
= &\iint_{\chi_n^{(t)}, \chi_n^{(t+1)}}\tilde p(\chi_n^{(t)},\chi_n^{(t+1)}|\boldsymbol \phi_n^{(t)})\log \frac{\tilde p(\chi_n^{(t)}|\boldsymbol \phi_n^{(t)})p(\chi_n^{(t+1)}|\chi_n^{(t)})}{q(\chi_n^{(t)}|\boldsymbol \phi_n^{(t)})p(\chi_n^{(t+1)}|\chi_n^{(t)})}\\
= &\iint_{\chi_n^{(t)}, \chi_n^{(t+1)}}\tilde p(\chi_n^{(t)},\chi_n^{(t+1)}|\boldsymbol \phi_n^{(t)})\left(\log \frac{\tilde p(\chi_n^{(t+1)}|\boldsymbol \phi_n^{(t)})}{q(\chi_n^{(t+1)}|\boldsymbol \phi_n^{(t)})}+\log \frac{\tilde p(\chi_n^{(t)}|\chi_n^{(t+1)},\boldsymbol \phi_n^{(t)})}{ q(\chi_n^{(t)}|\chi_n^{(t+1)},\boldsymbol \phi_n^{(t)})}\right)\\
= &\mathbb D[\tilde p(\chi_n^{(t+1)}|\boldsymbol \phi_n^{(t)})||q(\chi_n^{(t+1)}|\boldsymbol \phi_n^{(t)})]+\int_{\chi_n^{(t+1)}}\tilde p(\chi_n^{(t+1)}|\boldsymbol \phi_n^{(t)})\mathbb D[\tilde p(\chi_n^{(t)}|\chi_n^{(t+1)},\boldsymbol \phi_n^{(t)})|| q(\chi_n^{(t)}|\chi_n^{(t+1)},\boldsymbol \phi_n^{(t)})]\\
\ge &\mathbb D[\tilde p(\chi_n^{(t+1)}|\boldsymbol \phi_n^{(t)})||q(\chi_n^{(t+1)}|\boldsymbol \phi_n^{(t)})].
\end{split}
\end{equation}
Note that the integration will be replaced by the summation if dealing with $a_n^{(t)}$. For concise, we omit the proof of the second inequality, which can be proved in the same way. As a consequence, we obtain the (\ref{propo1}).

\bibliography{a}

\begin{thebibliography}{10}
\providecommand{\url}[1]{#1}
\csname url@samestyle\endcsname
\providecommand{\newblock}{\relax}
\providecommand{\bibinfo}[2]{#2}
\providecommand{\BIBentrySTDinterwordspacing}{\spaceskip=0pt\relax}
\providecommand{\BIBentryALTinterwordstretchfactor}{4}
\providecommand{\BIBentryALTinterwordspacing}{\spaceskip=\fontdimen2\font plus
\BIBentryALTinterwordstretchfactor\fontdimen3\font minus
  \fontdimen4\font\relax}
\providecommand{\BIBforeignlanguage}[2]{{%
\expandafter\ifx\csname l@#1\endcsname\relax
\typeout{** WARNING: IEEEtran.bst: No hyphenation pattern has been}%
\typeout{** loaded for the language `#1'. Using the pattern for}%
\typeout{** the default language instead.}%
\else
\language=\csname l@#1\endcsname
\fi
#2}}
\providecommand{\BIBdecl}{\relax}
\BIBdecl

\bibitem{YuOn2017}
W.~{Yu}, ``On the fundamental limits of massive connectivity,'' in \emph{Proc.
  Inf. Theory Appl. Workshop}, Feb. 2017, pp. 1--6.

\bibitem{BockelmannMassive2016}
C.~{Bockelmann}, N.~{Pratas}, and H.~{Nikopour}, ``Massive machine-type
  communications in 5{G}: physical and {MAC}-layer solutions,'' \emph{IEEE
  Commun. Mag.}, vol.~54, no.~9, pp. 59--65, Sep. 2016.

\bibitem{LiuMassive12018}
L.~{Liu} and W.~{Yu}, ``Massive connectivity with massive {MIMO}-part {I}:
  Device activity detection and channel estimation,'' \emph{IEEE Trans. Signal
  Process.}, vol.~66, no.~11, pp. 2933--2946, Jun. 2018.

\bibitem{LiuSparse2018}
L.~Liu, E.~G. Larsson, W.~Yu, P.~Petar, and \emph{et al}., ``Sparse signal
  processing for grant-free massive connectivity: A future paradigm for random
  access protocols in the internet of things,'' \emph{IEEE Signal Process.
  Mag.}, vol.~35, no.~5, pp. 88--99, Sep. 2018.

\bibitem{SenelGrant-Free2018}
K.~{Senel} and E.~G. {Larsson}, ``Grant-free massive {MTC}-enabled massive
  {MIMO}: A compressive sensing approach,'' \emph{IEEE Trans. Commun.},
  vol.~66, no.~12, pp. 6164--6175, Dec. 2018.

\bibitem{Schepker2013Exploiting}
H.~F. {Schepker}, C.~{Bockelmann}, and A.~{Dekorsy}, ``Exploiting sparsity in
  channel and data estimation for sporadic multi-user communication,'' in
  \emph{Proc. Int. Symp. Wireless Commun. Syst.}, Aug 2013, pp. 1--5.

\bibitem{Wunder2015Compressive}
G.~{Wunder}, P.~{Jung}, and M.~{Ramadan}, ``Compressive random access using a
  common overloaded control channel,'' in \emph{Proc. IEEE Globecom Workshops},
  Dec 2015, pp. 1--6.

\bibitem{Xu2015Active}
X.~{Xu}, X.~{Rao}, and V.~K.~N. {Lau}, ``Active user detection and channel
  estimation in uplink cran systems,'' in \emph{Proc. IEEE Int. Conf. Commun.
  (ICC)}, June 2015, pp. 2727--2732.

\bibitem{DonohoMessage}
D.~L. Donoho, A.~Maleki, and A.~Montanari, ``Message-passing algorithms for
  compressed sensing,'' \emph{Proc. Nat. Acad. Sci. {USA}}, vol. 106, no.~45,
  pp. 18\,914--18\,919, Nov.

\bibitem{ChenSparse2018}
Z.~{Chen}, F.~{Sohrabi}, and W.~{Yu}, ``Sparse activity detection for massive
  connectivity,'' \emph{IEEE Trans. Signal Process.}, vol.~66, no.~7, pp.
  1890--1904, Apr. 2018.

\bibitem{HannakJoint2015}
G.~{Hannak}, M.~{Mayer}, A.~{Jung}, G.~{Matz}, and N.~{Goertz}, ``Joint channel
  estimation and activity detection for multiuser communication systems,'' in
  \emph{Prob. IEEE Int. Conf. Commun. Workshop (ICCW)}, Jun. 2015, pp.
  2086--2091.

\bibitem{YangWireless2018}
Q.~{Yang}, H.~M. {Wang}, T.~X. {Zheng}, Z.~{Han}, and M.~H. {Lee}, ``Wireless
  powered asynchronous backscatter networks with sporadic short packets:
  Performance analysis and optimization,'' \emph{IEEE Internet Things J.},
  vol.~5, no.~2, pp. 984--997, Apr. 2018.

\bibitem{Dynamic2016}
B.~{Wang}, L.~{Dai}, Y.~{Zhang}, T.~{Mir}, and J.~{Li}, ``Dynamic compressive
  sensing-based multi-user detection for uplink grant-free noma,'' \emph{IEEE
  Commun. Lett.}, vol.~20, no.~11, pp. 2320--2323, 2016.

\bibitem{Efficient2017}
Y.~{Du}, B.~{Dong}, Z.~{Chen}, X.~{Wang}, Z.~{Liu}, P.~{Gao}, and S.~{Li},
  ``Efficient multi-user detection for uplink grant-free noma:
  Prior-information aided adaptive compressive sensing perspective,''
  \emph{IEEE J. Sel. Areas Commun.}, vol.~35, no.~12, pp. 2812--2828, 2017.

\bibitem{Angelosante20091}
D.~{Angelosante}, G.~B. {Giannakis}, and E.~{Grossi}, ``Compressed sensing of
  time-varying signals,'' in \emph{Proc. Int. Conf. Digit. Signal Process.},
  2009, pp. 1--8.

\bibitem{Vaswani2010}
N.~{Vaswani} and W.~{Lu}, ``{Modified-CS}: Modifying compressive sensing for
  problems with partially known support,'' \emph{IEEE Trans. Signal Process.},
  vol.~58, no.~9, pp. 4595--4607, 2010.

\bibitem{Angelosante20092}
D.~{Angelosante}, S.~I. {Roumeliotis}, and G.~B. {Giannakis}, ``{Lasso-Kalman}
  smoother for tracking sparse signals,'' in \emph{Proc. Asilomar Conf.
  Signals, Syst. Comput.}, 2009, pp. 181--185.

\bibitem{ZinielDynamic2013}
J.~{Ziniel} and P.~{Schniter}, ``Dynamic compressive sensing of time-varying
  signals via approximate message passing,'' \emph{IEEE Trans. Signal
  Process.}, vol.~61, no.~21, pp. 5270--5284, Nov. 2013.

\bibitem{MaSparse2019}
J.~{Ma}, S.~{Zhang}, H.~{Li}, F.~{Gao}, and S.~{Jin}, ``Sparse bayesian
  learning for the time-varying massive {MIMO} channels: Acquisition and
  tracking,'' \emph{IEEE Trans. Commun.}, vol.~67, no.~3, pp. 1925--1938, Mar.
  2019.

\bibitem{PrasadJoint2015}
R.~{Prasad}, C.~R. {Murthy}, and B.~D. {Rao}, ``Joint channel estimation and
  data detection in {MIMO}-{OFDM} systems: A sparse bayesian learning
  approach,'' \emph{IEEE Trans. Signal Process.}, vol.~63, no.~20, pp.
  5369--5382, Oct. 2015.

\bibitem{ZinielEfficient2013}
J.~{Ziniel} and P.~{Schniter}, ``Efficient high-dimensional inference in the
  multiple measurement vector problem,'' \emph{IEEE Trans. Signal. Process.},
  vol.~61, no.~2, pp. 340--354, Jan. 2013.

\bibitem{Rush2016}
C.~{Rush} and R.~{Venkataramanan}, ``Finite-sample analysis of approximate
  message passing,'' in \emph{Proc. IEEE ISIT}, 2016, pp. 755--759.

\bibitem{SchniterTurbo2010}
P.~{Schniter}, ``Turbo reconstruction of structured sparse signals,'' in
  \emph{Prob. Conf. Inf. Sci. and Syst. (CISS)}, Mar. 2010, pp. 1--6.

\bibitem{KschischangFactor2001}
F.~R. {Kschischang}, B.~J. {Frey}, and H.~. {Loeliger}, ``Factor graphs and the
  sum-product algorithm,'' \emph{IEEE Trans. Inf. Theory}, vol.~47, no.~2, pp.
  498--519, Feb. 2001.

\bibitem{BayatiDynamics2011}
M.~{Bayati} and A.~{Montanari}, ``The dynamics of message passing on dense
  graphs, with applications to compressed sensing,'' \emph{IEEE Trans. Inf.
  Theory}, vol.~57, no.~2, pp. 764--785, Feb. 2011.

\bibitem{Boyen2013}
X.~{Boyen} and D.~{Koller}, ``Tractable inference for complex stochastic
  processes,'' 2013, [Online]. Available: https://arxiv.org/abs/1301.7362,
  preprint.

\bibitem{cover2005}
T.~M. C. J.~A. Thomas, \emph{Elements of Information Theory}.\hskip 1em plus
  0.5em minus 0.4em\relax Wiley-Interscience, 2005.

\bibitem{Burnham1998Model}
K.~P. Burnham and D.~R. Anderson, ``Model selection and inference. a practical
  information-theoric approach,'' \emph{Technometrics}, vol.~45, no.~2, pp.
  181--181, 1998.

\bibitem{Opper1999}
M.~{Opper} and O.~{Winther}, \emph{A Bayesian approach to on-line learning},
  Jan. 1999.

\bibitem{Montanari2011}
A.~{Montanari}, ``Graphical models concepts in compressed sensing,'' 2011,
  [Online]. Available: https://arxiv.org/abs/1011.4328, preprint.

\bibitem{TreesDetection2001}
H.~L. Van~Trees, \emph{Detection, Estimation, and Modulation Theory, Part I :
  Detection, Estimation, and Linear Modulation Theory.}\hskip 1em plus 0.5em
  minus 0.4em\relax Wiley-Interscience, 2001.

\bibitem{OMP2007}
J.~A. {Tropp} and A.~C. {Gilbert}, ``Signal recovery from random measurements
  via orthogonal matching pursuit,'' \emph{IEEE Trans. Inf. Theory}, vol.~53,
  no.~12, pp. 4655--4666, 2007.

\bibitem{DonohoMessage2010}
D.~L. {Donoho}, A.~{Maleki}, and A.~{Montanari}, ``Message passing algorithms
  for compressed sensing: I. motivation and construction,'' in \emph{Proc. IEEE
  Inf. Theory Workshop}, Jan. 2010, pp. 1--5.

\end{thebibliography}

\end{document}